\documentclass[aps,pra,twocolumn]{revtex4}
\usepackage[dvips]{epsfig}
\usepackage[usenames]{color}
\usepackage{pstricks}
\usepackage{amsmath}
\usepackage{amssymb}
\usepackage{subfigure}
\usepackage{afterpage}
\usepackage{multirow}
\begin{document}

\title{Universalities in ultracold reactions of alkali polar molecules}

\author{Goulven Qu{\'e}m{\'e}ner}
\email{goulven.quemener@colorado.edu}
\author{John L. Bohn}
\affiliation{JILA, University of Colorado,
Boulder, C0 80309-0440, USA}

\author{Alexander Petrov}
\altaffiliation{Alternative address: St. Petersburg Nuclear Physics Institute, 
Gatchina, 188300; St.Petersburg State University, 198904, Russia}
\author{Svetlana Kotochigova}
\affiliation{Temple University, Philadelphia, PA 19122, USA}

\date{\today}

\begin{abstract}

We consider ultracold collisions of ground-state, heteronuclear alkali
dimers that are susceptible to four-center chemical reactions 2~AB
$\to$ A$_2$ + B$_2$ even at sub-microKelvin temperature.
These reactions depend strongly on species, temperature, electric
field, and confinement in an optical lattice.  We calculate {\it ab
initio} van der Walls coefficients for these interactions, 
and use a quantum formalism to study the scattering properties of such molecules
under an external electric field and optical lattice. 
We also apply a quantum threshold model to explore the dependence of reaction rates on
the various parameters.  We find that, among the heteronuclear alkali fermionic species, LiNa is
the least reactive, whereas LiCs is the most reactive.
For the bosonic species, LiK is the most reactive in zero field, but all
species considered  – LiNa, LiK, LiRb, LiCs, and KRb – share a
universal reaction rate once a sufficiently high electric field is
applied.
\end{abstract}

%\pacs{ }

\maketitle

\font\smallfont=cmr7

\section{Introduction}

The study of ultracold polar molecules has now become 
a vast and exciting area of interest 
since the formation of bi-alkali 
heteronuclear polar molecules~\cite{Sage05,Hudson08,Ni08,Deiglmayr08,Aikawa10}.
The molecules can be controlled at 
the ground electronic, vibrational, rotational~\cite{Ni08}, 
and hyperfine~\cite{Ospelkaus10-PRL}
quantum-level.
The external motion of the polar molecules
can also be modified by an electric field~\cite{Ni10-NATURE} 
and by an optical lattice confinement~\cite{Miranda11}.

Polar molecules offer remarkable characteristics.
First, they have strong electric dipole moments~\cite{Aymar05,Deiglmayr08-JCP}. 
The interactions between polar molecules
can then be dominated by electric dipole-dipole terms.
The electric molecular interactions
are strong, long-range, anisotropic and can be tuned by electric fields. 
Secondly, the polar molecules can be either bosons or fermions.
If the polar molecules are addressed in a single quantum state,
they become indistinguishable and quantum statistics
plays a strong role. An ultracold gas of bosonic molecules
can lead to Bose-Einstein condensation
and an ultracold sample of fermionic molecules can lead
to a Degenerate Fermi gas.
Thirdly, two polar molecules can be reactive or not~\cite{Zuchowski10,Byrd10,Meyer10,Meyer11}.
It was found in Ref.~\cite{Zuchowski10}
that among the bi-alkali heteronuclear molecules in their absolute fundamental 
ground state, that 
the Lithium species LiNa, LiK, LiRb, LiCs in addition with 
the KRb molecule (category 1) gave rise to two-body exoergic chemical reactive processes while
the remaining species NaK, NaRb, NaCs, KCs, RbCs (category 2) resulted 
in two-body endoergic processes.
Reactivity is an advantage to investigate the ultracold chemistry
of molecules~\cite{Ospelkaus10-SCIENCE}. 
It also provides a clear signature (in term of molecular loss)
of two-body interactions
in a gas and depends strongly on the applied electric field~\cite{Quemener10-QT}.
The non-reactive molecules have the advantage
of being chemically stable in their absolute ground state
and can help to reach long-lived samples of polar molecules.
However, 
if dense samples of molecules are formed in Bose-Einstein condensates for example,
three-body collision can become a source of loss, and it is important to investigate 
the collisional properties of such 
processes~\cite{Ticknor10-3B,Wang11-BOS,Wang11-FER}.
Finally, molecules offer a rich internal quantum structure
and can be manipulated with electromagnetic waves
in order to address their quantum state.
Exciting perspectives have been proposed for these polar molecules.
This involves condensed matter and many-body physics, quantum magnetism, precision measurements,
controlled chemistry and 
quantum information~\cite{Carr09,Micheli06-NATURE,Pupillo-Chapter,Demille02,Yelin06,Gorshkov11-PRL,Gorshkov11-PRA}.

For all these reasons, many experimental groups are currently interested
in creating polar molecules.
The fermionic polar molecules $^{40}$K$^{87}$Rb received a particular
experimental~\cite{Ospelkaus08,Ni08,Ospelkaus10-PRL,Ospelkaus10-SCIENCE,Ni10-NATURE,Miranda11} and 
theoretical~\cite{Kotochigova09,Quemener10-QT,Idziaszek10-PRL,Ticknor10,Kotochigova10,Quemener10-2D,Micheli10-PRL,Idziaszek10-RAPID,Gao10,Quemener11-FULL,Dincao10,Julienne11-PCCP}
consideration recently.
However, much less is known about the interactions and the dynamical properties of the 
other polar bi-alkali molecules, for which experimental attention is also
devoted~\cite{Sage05,Hudson08,Haimberger09,Zabawa10,Lercher11,Debatin11,Cho11,Deiglmayr08,Deiglmayr11-JPCS,Deiglmayr11-EPJD,Ridinger11}.
This is what we address in this article. In Section II, 
we compute the isotropic long-range van der Waals coefficients between polar molecules.
We focus our study to the exoergic molecules (category 1). 
In Section III, we use these parameters to perform quantum scattering calculations assuming full loss
when the polar molecules are close to each other.
We consider the case of collisions in free and confined space, in electric fields.
We use a Quantum Threshold (QT) model to explain how the collisional properties scale
with the different species.
We arrive at analytical expressions of high-loss collision rates of bosonic or fermionic molecules,
which can also be applied to the inelastic and reactive case of 
molecules of category 2, as well as atom-atom or atom-molecule collisions,
provided the van der Waals coefficients are known.
We conclude in Section IV.

\section{Isotropic long-range interaction of reactive polar molecules}

The isotropic dispersion coefficient $C_6$ between two identical diatomic
alkali-metal molecules in the $v$=0 and $J$=0 rovibrational ground state
of the X$^1\Sigma^+$ potential has three contributions
\begin{eqnarray}
C_6 &=& C_6^{(\rm gr)}+ C_6^{(\rm exc)} + C_6^{(\rm inf)} \label{C6}\\ 
        &=& \frac{3}{\pi}\int_0^{\infty}d\omega \left\{\alpha^2_{\rm gr}(i\omega) + \alpha^2_{\rm exc}(i\omega)\right. \nonumber \\
        & &  \quad \quad\quad \quad\quad\quad \quad\quad
                 \left. + 2\alpha_{\rm gr}(i\omega)\alpha_{\rm exc}(i\omega)\right\}\,,
\nonumber
\end{eqnarray}
where the first term of the integrand is the square of the isotropic
dynamic polarizability $\alpha_{\rm gr}(i\omega)$ at imaginary frequency
$i\omega$ due to  rovibrational transitions within the ground state
potential. The second term in the integrand is the square of the
isotropic polarizability $\alpha_{\rm exc}(i\omega)$ due to transitions
to the rovibrational levels of electronically excited potentials, while the last term indicates an
interference between the first two contributions.  In these and subsequent
expressions both the dispersion coefficient and the polarizability are
in atomic units.  A thorough discussion of dispersion forces between
molecules can be found in Ref.~\cite{Stone}.

We find that  $\alpha_{\rm gr}(i\omega)=\alpha_{0g}/(1+(\omega/\eta_g)^2)$
\cite{Stone,Tang} to good approximation with $\alpha_{0g} = d_\text{p}^2/(3B)$
and $\eta_{g} = 2B$, where $d_\text{p}$ and $B$ are the electric permanent dipole moment and
rotational constant at the equilibrium separation $R_e$ between the atoms
in the molecule, respectively. The contribution from transitions between
vibrational levels within the ground state potential is negligibly small.
Consequently, $C_6^{(\rm gr)}={d_\text{p}^4}/(6B)$ in agreement with the findings of Ref.~\cite{Barnett}.

The isotropic dynamic polarizability $\alpha_{\rm exc}(i\omega)$ contains
contributions from transitions to the rovibrational excited $^1\Sigma^+$ and $^1\Pi$
potentials, which correspond to the parallel and perpendicular component
of the polarizability, respectively.  Based on the Franck-Condon principle
we can evaluate the polarizability at each interatomic separation rather
than perform an average over ro-vibrational levels \cite{Herzberg}. For
$v$=0 and $J$=0 the separation is $R=R_e$.  We then parametrize
$\alpha_{\rm exc}(i\omega)=\sum_j \alpha_j(i\omega)$ with 
\begin{equation}
\alpha_j(i\omega)= \frac{\alpha_{0j}}{1+(\omega/\eta_j)^2} \,.
\label{polar} 
\end{equation} 
Each term corresponds to an excited potential.  In practice, we have
found it more convenient to evaluate the polarizability at $R=R_e$
as function of real frequencies and find the parameters $\alpha_{0j}$
and $\eta_j$ from a fit.  The static polarization due to the excited
state potentials is $\alpha_{\rm exc}(0)=\sum_j \alpha_{0j}$.
Using Eqs.~(\ref{C6}) and (\ref{polar}) we obtain  
the $C_6^{(\rm exc)}$ and $C_6^{(\rm inf)}$ coefficients as 
\begin{eqnarray}
 C_6^{(\rm exc)} & =& \frac{3}{2} \sum_{jk}
 \frac{\alpha_{0j}\alpha_{0k}}{1/\eta_j+1/\eta_k} \\ C_6^{(\rm inf)}
 &=& 3 \sum_{j}  \frac{\alpha_{0g}\alpha_{0j}}{1/\eta_g+1/\eta_j} \,.
\end{eqnarray} 

The dynamic polarizability $\alpha_{\rm exc}(\omega)$ at real
frequency is calculated using a coupled cluster method with single
and double excitations (ccsd) \cite{ccsd}.  The calculation of the
static polarizability and permanent dipole moment is performed at
much higher level using coupled cluster method with the single,
double and triple excitations (ccsdt).  Twelve electrons, including
$1s^22s^1$ of the Li atom and $(n-1)s^2(n-1)p^6ns^1$ of the Na,
K, Rb, and Cs atoms, were explicitly used in both ccsd and ccsdt
calculations.  The dipole moment for each molecule was averaged on the
zero vibrational level.  We employed the cc-pCVQZ basis sets for Li and
Na from Refs.~\cite{Dunning:89,Prascher:11}, the all-electron basis
for the K atom from Ref.~\cite{KPeterson2009}, and the  ECP28MDF and
ECP46MDF basis sets with the relativistic effective core potentials
from Ref.~\cite{Stoll} for the Rb and Cs atoms.  A comparison of our
data on the dipole moment and static polarizability with results of
Refs.~\cite{Aymar05,Deiglmayr08-JCP} shows a good agreement within a few \%.

Table~\ref{data} lists our $C_6$ coefficients  for four pairs of identical
alkali-metal molecules in the $v=0$, $J=0$  rovibrational level of the X
$^1\Sigma^+$ potential.  For completeness, we tabulate the contribution to
the isotropic component of the static polarizability from electronically
excited potentials, the rotational constant, and the permanent dipole
moment for each of the four molecules.

\begin{table}
\caption{Van der Waals $C_6$ coefficients in atomic units for the
interaction between the two molecules in the $v=0$, $J=0$ rovibrational
levels of the X $^1\Sigma^+$ potential and other molecular characteristics
used to calculate $C_6$; $\alpha_{\rm exc}(0)$  is the isotropic static
polarizability due to transitions to electronically excited potentials;
$B$ and $d_\text{p}$ are the rotational constant and electric permanent dipole moment,
respectively. These three properties are evaluated at the equilibrium
separation $R_e$.  The value of $B$ is from \cite{Deiglmayr08-JCP}. The next
three columns are the excited state, interference, and ground state
contributions to the total $C_6$, shown in the last column.
} \label{data}
\vspace*{0.2cm}
\begin{tabular}{lllrrrr} \hline
$\alpha_{\rm exc}(0)$~~~ & $B/hc$~~~  & ~~$d_\text{p}$~ & ~~$C_6^{({\rm exc})}$~ &
~~$C_6^{({\rm inf})}$~ & ~~$C_6^{({\rm gr})}$~& $C_6$~~~\\
(a.u.) &  (cm$^{-1}$)  & (D) & (a.u.) & (a.u.) & (a.u.) & (a.u.)\\\hline
\multicolumn{7}{c}{ LiNa + LiNa} \\
\hline
237.8 &  0.377 & 0.557$^a$ & 3673 &  23 &    222 &   3917 \\
      &  & 0.531$^b$ & 3673 &  21 &    186 &   3880 \\ \hline
\multicolumn{7}{c}{ LiK + LiK}\\ \hline
324.9 &   0.258 & 3.556$^a$ & 6269 &  1271 &   542000&   550000  \\
      &         & 3.513$^b$ & 6269 &  1241 &   517000 &  524000   \\ \hline
\multicolumn{7}{c}{ LiRb + LiRb} \\ \hline
346.2 &   0.220& 4.130$^a$ &   6323  & 1829 &   1160000 & 1170000 \\
      &        & 4.046$^b$ &   6323  & 1754  &  1070000 & 1070000 \\ \hline
\multicolumn{7}{c}{ LiCs + LiCs} \\ \hline
389.7 &   0.188 & 5.478$^a$ &  7712 & 3620  &   4200000 & 4210000 \\
      &         & 5.355$^b$ & 7712  & 3460  &   3830000 & 3840000
\end{tabular}
\footnotetext{$^a$ Ref. \cite{Aymar05}}
\footnotetext{$^b$ This work}
\end{table}

Table \ref{data} shows that the value of the $C_6$ coefficient as well
as the three contributions to it  increase when we move down along
the first column of the periodic table for the second atom in our
four diatomic molecules. Most of the increase can be traced back to
increasing permanent and transition dipole moments.  For example, the
ground state contribution $C_6^{({\rm gr})}$ increase by four order of
magnitude as the permanent dipole moment increases by a factor of ten.
For the excited state contribution $C_6^{({\rm exe})}$ the increase is
less dramatic as the transition dipole moments increase only weakly. Only
for the LiNa molecule does the excited state contribution dominate the
$C_6$ coefficient.

\section{Dynamics in three-dimensional space}

\subsection{Quantum numerical calculation}

We use the isotropic van der waals $C_6$ coefficients
calculated in the previous section to compute
the chemical rate coefficients
of the reactive polar molecules.
We use the same formalism used in Ref.~\cite{Ni10-NATURE}
for the chemical reaction KRb + KRb $\to$ K$_2$ + Rb$_2$.
We employ a time-independent quantum formalism, including only one molecule-molecule channel
corresponding to the initial state of the molecules, but including several partial waves.
For two particles of mass $m_1,m_2$, the Hamiltonian of the system is given by
\begin{eqnarray}
H =  T + V_{\text{abs}}  
+ V_{\text{vdW}}  
+  V_{\text{dd}}.
\label{Hamiltonian}
\end{eqnarray}
Using spherical coordinates $(r,\theta,\phi)$, the kinetic energy is 
$T = - \hbar^2 \nabla^2_{\vec{R}} /(2 \mu)$ ,
$\mu = m_1 m_2 / (m_1 + m_2)$ is the reduced mass of the colliding system,
$ V_{\text{abs}} = i A e^{-(r-r_\text{min})/r_\text{c}}$ is an absorbing potential
to account for the loss of particles due to chemical reactions or inelastic collisions in the incident channel,
where $A$ is the strength of the absorbing potential, $r_\text{min}$ is the position where the potential starts,
and $r_\text{c}$ is the position where the potential vanishes exponentially. 
$V_{\text{vdW}} =   - C_6/R^6 $ is an isotropic van der Waals interaction, 
and $V_{\text{dd}} =  [d_1 \, d_2 \, (1 - 3 \cos^2{\theta})] / (4 \pi \varepsilon_0 \, R^3)$
is the dipole-dipole interaction between the two particles if an electric field is applied.
Here $d_1,d_2$ are the induced electric dipole moments in the laboratory frame 
and their maximum value is given by their
permanent dipole moment $d_\text{p,1},d_\text{p,2}$ in the molecular frame.
We expand the total wavefunction onto a basis set of spherical harmonics (or partial waves)
\begin{eqnarray}
\Psi^{M_L}(R,\theta,\varphi) =  \frac{1}{R} \, \sum_{L'}  Y^{M_L}_{L'}(R,\theta) 
\,  F^{M_L}_{L'}(R),
\label{Psisph}
\end{eqnarray}
where $L$ is the quantum number associated
with the orbital angular momentum of the collision,
and $M_L$, the quantum number associated with its projection
onto a quantization axis (see Ref.~\cite{Quemener10-QT} for details).
Solving the eigenstates of the Hamiltonian leads 
to the set of close-coupling equations
\begin{multline}
 \left\{ - \frac{\hbar^2}{2 \mu} \frac{d^2}{d R^2} 
 + V_\text{eff} + V_\text{abs} - E \right\}
\, F^{M_L}_{L L}(R) \\
+ \sum_{L' \ne L } 
  - \frac{C_3(L,L';M_L)}{R^3}
\  F^{M_L}_{LL'}(R) = 0 .
\label{eqcoup}
\end{multline}
$E$ represents the total energy which is, in this study, the collision energy $E_c$, 
as we use only one molecule-molecule incident channel.
We use the same notation as in Ref.~\cite{Quemener10-QT}
$C_3(L,L';M_L) = \alpha_{L,L'}^{M_L} \, d_1 \, d_2 \, / 4 \pi \varepsilon_0$
with
\begin{multline}
\alpha_{L,L'}^{M_L} = 2 \, (-1)^{M_L} \, \sqrt{2L+1} \, \sqrt{2L'+1} \\
\left( \begin{array}{ccc} L & 2 & L'  \\ 0 & 0 & 0 \end{array} \right) \,
\left( \begin{array}{ccc} L & 2 & L'  \\ - \, M_L & 0 & M_L'  \end{array} \right) \, \delta_{M_L,M_L'}.
\label{C3}
\end{multline}
The effective potential in Eq.~\eqref{eqcoup} is given by
\begin{eqnarray}
V_\text{eff} =  \frac{\hbar^2 \, L (L+1)}{2 \mu R^2}  - \frac{C_6}{R^6} - \frac{C_3(L,L;M_L)}{R^3}
\end{eqnarray}
for a given $L,M_L$.
The absorbing potential is chosen in Eq.~\eqref{eqcoup} in such a way that 
the elastic probability vanishes (or the loss probability is unity)
when the two molecules come close together.
The case for which the loss probability is smaller than unity has been discussed
in Ref.~\cite{Idziaszek10-RAPID,Idziaszek10-PRL,Kotochigova10}.

\begin{figure} [t]
\begin{center}
\includegraphics*[width=7.5cm,keepaspectratio=true,angle=0]{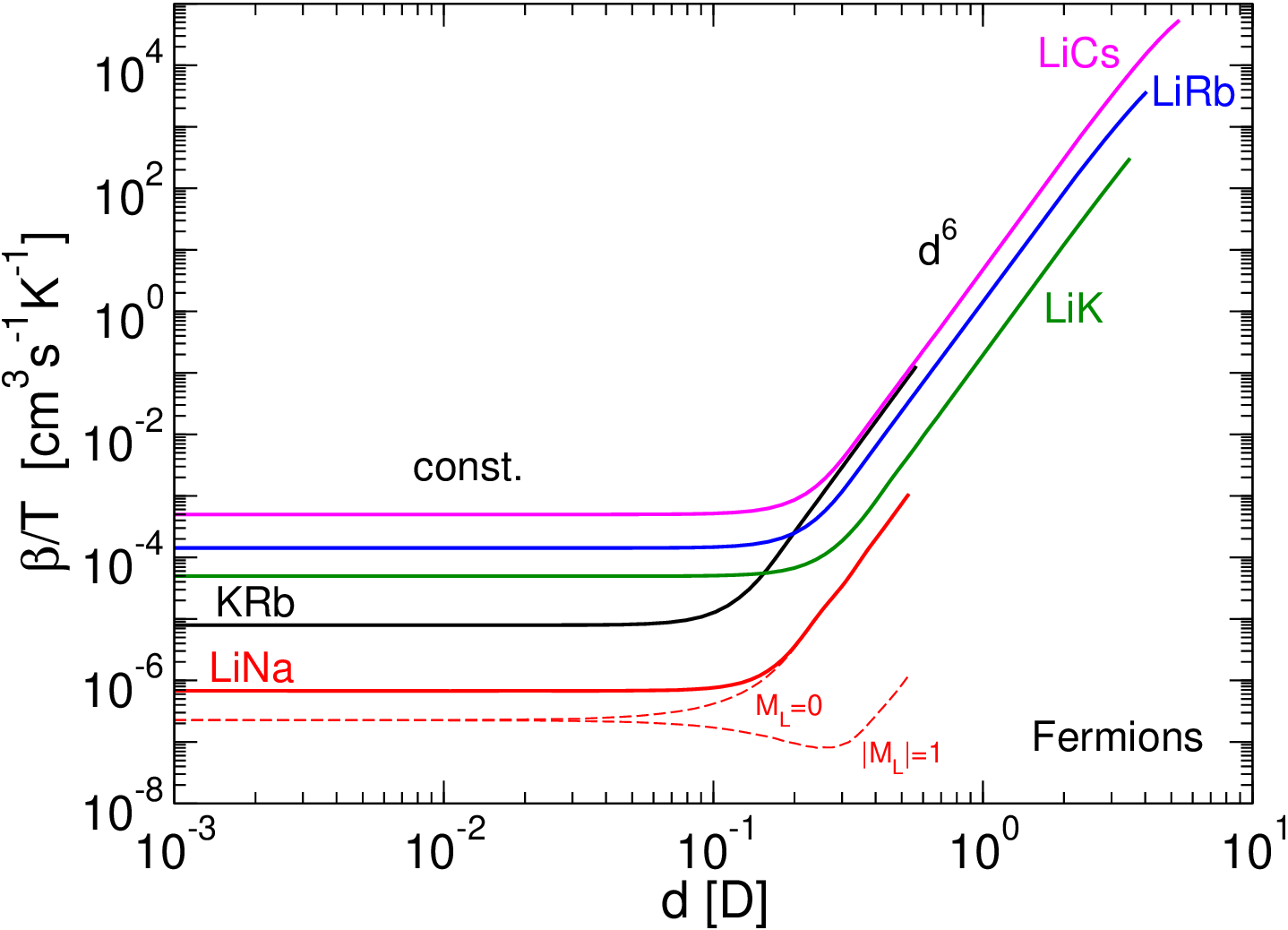}  
\caption{(Color online) 
Solid lines: Loss rate coefficient $\beta_{L=1}$ divided by $T$ in free 3D space, 
of the reaction AB + AB $\to$ A$_2$ + B$_2$ for different 
reactive fermionic polar molecules AB = LiNa, KRb, LiK, LiRb, LiCs, 
as a function of the electric dipole moment.
The fermions are considered in a same indistinguishable quantum state.
Dashed lines: $\beta_{L=1,M_L=0}$ and $\beta_{L=1,|M_L|=1}$ components of $L=1$, shown here for AB = LiNa.
\label{FER3D-TOTAL-FIG}
}
\end{center}
\end{figure}

\begin{figure} [t]
\begin{center}
\includegraphics*[width=7.5cm,keepaspectratio=true,angle=0]{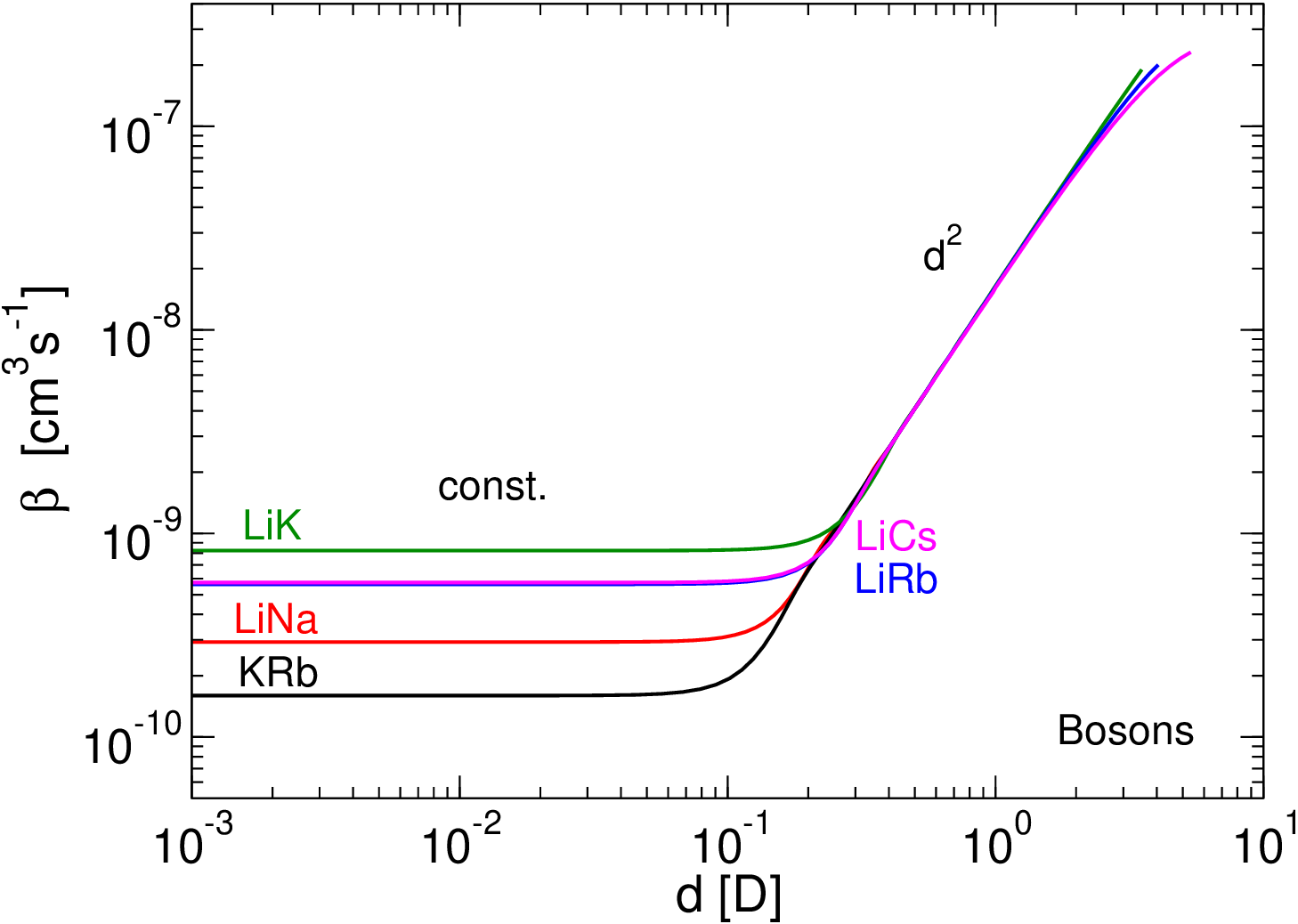}
\caption{(Color online) Same as Figure~\ref{FER3D-TOTAL-FIG} for different reactive bosonic polar molecules.
The rate coefficient $\beta_{L=0}$ is plotted as a function of the electric dipole moment for $T \to 0$ (Wigner regime).
The bosons are considered in a same indistinguishable quantum state.
\label{BOS3D-TOTAL-FIG}
}
\end{center}
\end{figure}

We report in Fig.~\ref{FER3D-TOTAL-FIG} and~\ref{BOS3D-TOTAL-FIG}
the loss rate coefficient 
as a function of the induced electric dipole moment $d$, 
for two indistinguishable fermionic molecules (Fig.~\ref{FER3D-TOTAL-FIG})
and for two indistinguishable bosonic molecules (Fig.~\ref{BOS3D-TOTAL-FIG}),
for LiNa--LiNa, KRb--KRb, LiK--LiK, LiRb--LiRb and LiCs--LiCs collisions.
To converge the results, we use five partial waves, $L=1,3,5,7,9$ for the fermions
and $L=0,2,4,6,8$ for the bosons.
We used the values of $C_6$ and $d_\text{p}$ reported in Tab.~\ref{data},
and the value of $C_6=16133$~a.u. of Ref.~\cite{Kotochigova10} and $d_p=0.566$~D of Ref.~\cite{Ni08} for KRb.
We provide a list of the fermionic and bosonic isotopes of each species in Appendix A.
These results have been obtained in the regime of ultracold temperature.
In this regime, the fermionic rate scales linearly with the temperature 
(hence we have plotted the rate divided by the temperature) 
while the bosonic rate is independent of the temperature according 
to the Bethe-Wigner laws~\cite{Bethe35,Wigner48}. 
For both cases, the rate scales as a constant in 
the van der Waals regime where $d \to 0$,
and an increasing term in the electric field regime 
where $d \to d_{\text{p}}$.
We note that for large dipole moments, the corresponding dipole length $a_\text{dd}= \mu d^2 / \hbar^2 $
may exceed the distance between molecules given by the inverse third of the molecular gas density $a_\text{mm}= n^{-1/3}$.
In such situation, there are no more collisions between molecules. Instead, a dense liquid/solid phase is entered
where many-body physics becomes important.

For the fermionic case, in the van der Waals regime,
it is seen that the LiNa system is the least reactive,
followed by KRb, LiK, LiRb and finally LiCs. Qualitativelly, 
light masses and small values of $C_6$ 
increase the incident p-wave barrier (this is the case for LiNa) and hence decrease 
the chance to get high chemical reactivity,
while heavy masses and large values of $C_6$ decrease the barrier (this is the case for LiCs)
and increase the reactivity.
In the electric field regime, 
the same general trend is observed, except now the rate of the KRb system
is as high as the LiCs system. Now the rates seem to scale with the reduced mass of the system only.
For a given dipole, the electric dipole interaction is the same between the species,
only the centrifugal terms differ. Higher mass means smaller barrier so higher loss rate.

For the bosonic case, in the van der Waals regime, KRb are the least reactive molecules, 
followed by LiNa, LiRb, LiCs and finally LiK.
Bosonic particles collide in a s-wave at ultralow energy where no incident barrier is present.
%Then, it becomes difficult to explain the behaviours qualitatively.
Instead, one must invoke the probability for quantum transmission.
In the electric field regime, all different systems have the same rate coefficients.
This will be explained in the next section.

\subsection{Quantum Threshold model}

To understand the physical trends seen in the numerical results, we employ 
an analytical Quantum Threshold model (QT model)~\cite{Quemener10-QT}
which provides a universal expression of 
an ultracold collision (chemical reaction or inelastic collision)
with short range unit loss probability.
The QT model is a clear and simple model
to describe the dependence of an ultracold chemical reaction
on the reduced mass and the isotropic van der waals $C_6$ coefficient
of the molecule-molecule complex, and on
the induced dipole moment via the presence of an applied electric field.
The QT model assumes that the loss probability scales as
\begin{eqnarray}
P_{L,M_L} = p_{_{L,|M_L|}} \, \bigg\{\frac{E_c}{E_*}\bigg\}^{L+1/2}
\label{QT}
\end{eqnarray}
where $E_*$ is a characteristic 
energy corresponding to the long range interaction 
of the molecules in a partial wave $L, M_L$.
$p_{_{L,|M_L|}}$ is a dimensionless quantity of order of unity, 
and is estimated by fitting the expression
with the numerical results. 
The thermalized rate coefficient is expressed by
\begin{eqnarray}
\beta_{L,M_L}  = p_{_{L,|M_L|}} \,
 \frac{\hbar^2  \pi}{\sqrt{2 \mu^3}} \ \frac{ \langle E_c^{L} \rangle } {E_*^{L+\frac{1}{2}} }   \times \Delta
\label{rateQT}
\end{eqnarray}
where the brackets denote a Maxwell-Boltzmann distribution over the collision energy to the power $L$.
$\Delta=2$ if the particles are in indistinguishable states
and $\Delta=1$ if they are in distinguishable states~\cite{Burke-PHD}.

\subsubsection{QT model for p-wave collisions}

For p-wave collision ($L=1$), 
we chose the characteristic energy $E_*$ equal to 
the height of the incident barrier, $E_{_{L,|M_L|}}^{n,m}$,
of the effective potential $V_\text{eff}$,
composed of the strongest attractive potential $-C_n / R^n$ 
and the strongest repulsive potential $C_m / R^m$
\begin{eqnarray}
E_{_{L,|M_L|}}^{n,m} = \frac{C_m \, \left(  \frac{n \, C_n}{m \, C_m} \right)^{\frac{n}{n-m}}  
- C_n \, \left(  \frac{n \, C_n}{m \, C_m} \right)^{\frac{m}{n-m}}  }{\left(  \frac{n \, C_n}{m \, C_m} \right)^{\frac{n+m}{n-m}}}.
\end{eqnarray}
The position of the barrier is given by
\begin{eqnarray}
R_{_{L,|M_L|}}^{n,m} = \left(  \frac{n \, C_n}{m \, C_m} \right)^{\frac{1}{n-m}}.
\end{eqnarray}
The combinations of $n$ and $m$ are given in Tab.~\ref{TAB2} with the corresponding height of the barriers.
For the van der Waals regime
and for either $|M_L|=0,1$, the height of the barrier is made by the 
the attractive van der Waals interaction $-C_6/R^6$ and
the repulsive centrifugal 
term $C_2/R^2 \equiv \hbar^2 \, L (L+1) / (2 \mu R^2)$ with $L=1$, giving rise to a 
characteristic energy $E_{{1,(0,1)}}^{6,2}$.
For the electric field regime
and for $|M_L|=0$, the height of the barrier is made by 
the attractive dipole-dipole interaction $-C_3(1,1;0)/R^3 \equiv - [(4/5) \, d^2 / (4 \pi \varepsilon_0)] / R^3$
and the repulsive centrifugal 
term $C_2/R^2 \equiv \hbar^2 \, L (L+1) / (2 \mu R^2)$ with $L=1$, giving rise to a 
characteristic energy $E_{{1,0}}^{3,2}$.
Finally, for the electric field regime
and for $|M_L|=1$, the height of the barrier is made by 
an attractive $-C_4/R^4 \equiv - [ (72 \, \mu /(875 \, \hbar^2)) \, d^4 / (4 \pi \varepsilon_0)^2] / R^4$
and the repulsive dipole-dipole interaction $-C_3(1,1;1)/R^3 \equiv +[(2/5) \, d^2 / (4 \pi \varepsilon_0)] / R^3$,
giving rise to a 
characteristic energy $E_{{1,1}}^{4,3}$.
The $-C_4/R^4$ attractive interaction comes from the coupling 
between the $L=1$ and $L=3$ of the $|M_L|=1$ component.
This is demonstrated in Appendix B.

\begin{table}[h]
\caption{Characteristic energies $E_*$ for $L=1, |M_L|=0,1$
in the van der Waals (vdW) and electric (elec.) regime.
\label{TAB2}
}
\begin{center}
\begin{tabular}{|c | c | c | c | c |}
\hline
regime & $|M_L|$ & - $C_n / R^n$ & $C_m / R^m$ & $E_{_{L=1,|M_L|}}^{n,m}$    \\ [0.5ex]
\hline
 vdW  &0,1 & - $\frac{C_6}{R^6}$ & $\frac{\hbar^2 \, L (L+1)}{2 \, \mu \, R^2}$ & $\left( \frac{8 \, \hbar^6}{54 \, \mu^3 \, C_6} \right)^{1/2}$ \\
 elec. &0   & - $\frac{ (4/5) \, d^2 }{ 4 \pi \varepsilon_0 \, R^3} $ & $ \frac{\hbar^2 \, L (L+1)}{2 \, \mu \, R^2}$ & $\frac{ 25 \, \hbar^6 }{ 108 \, \mu^3} \left( \frac{d^2}{4 \pi \varepsilon_0} \right)^{-2} $ \\
 elec. &1   & - $\frac{ 72 \, \mu \, d^4}{875 \, \hbar^2 \, (4 \pi \varepsilon_0)^2 \, R^4}$ & $\frac{ (2/5) \, d^2}{4 \pi \varepsilon_0 \, R^3} $ &  
$\frac{ (875/24)^3 \, \hbar^6 }{ 10000 \, \mu^3} \left( \frac{d^2}{4 \pi \varepsilon_0} \right)^{-2} $ \\ [0.5ex]
\hline
\end{tabular}
\end{center}
\end{table}

\begin{figure} [t]
\begin{center}
\includegraphics*[width=7.2cm,keepaspectratio=true,angle=0]{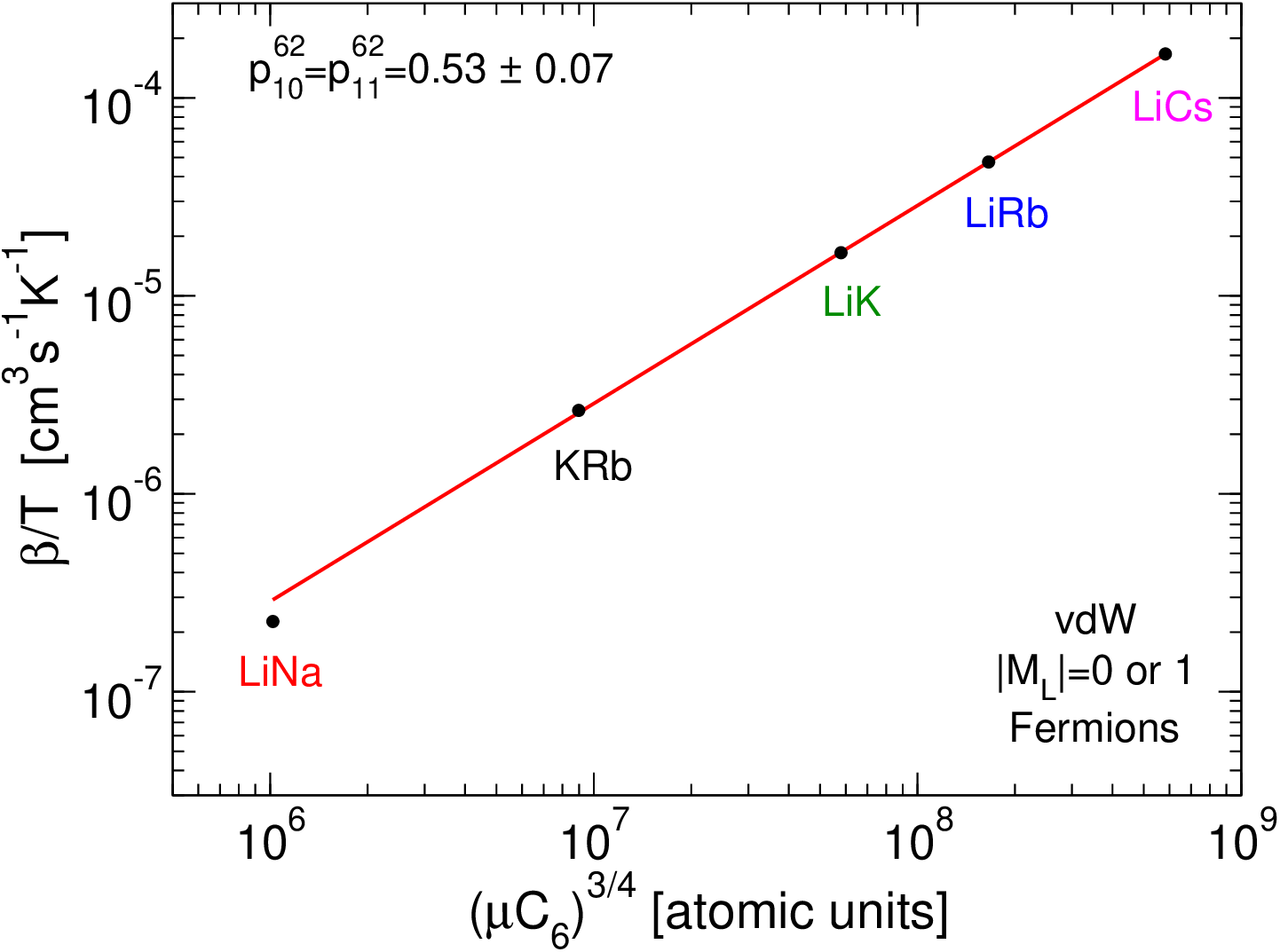}  \\
\includegraphics*[width=7.2cm,keepaspectratio=true,angle=0]{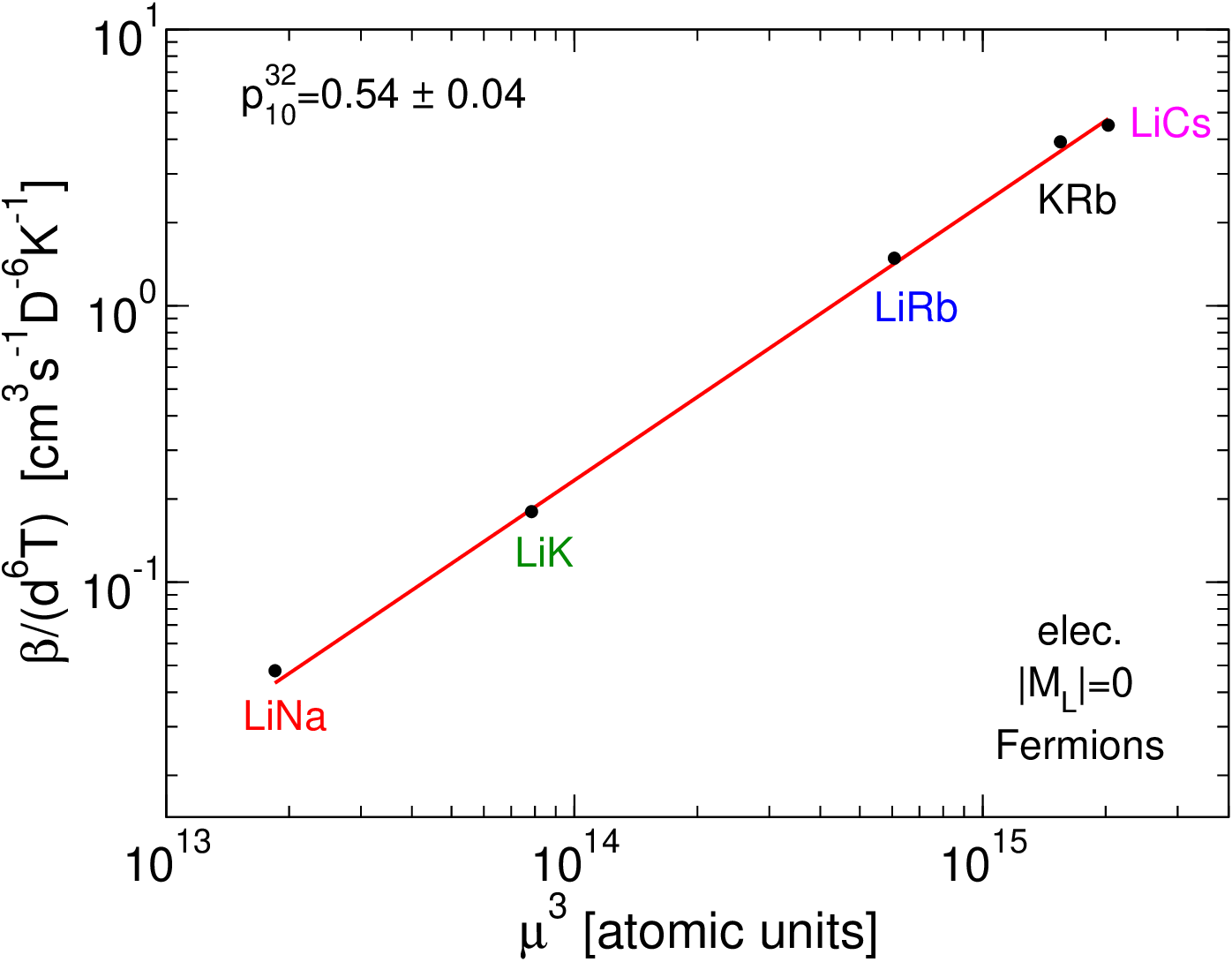}  \\
\includegraphics*[width=7.2cm,keepaspectratio=true,angle=0]{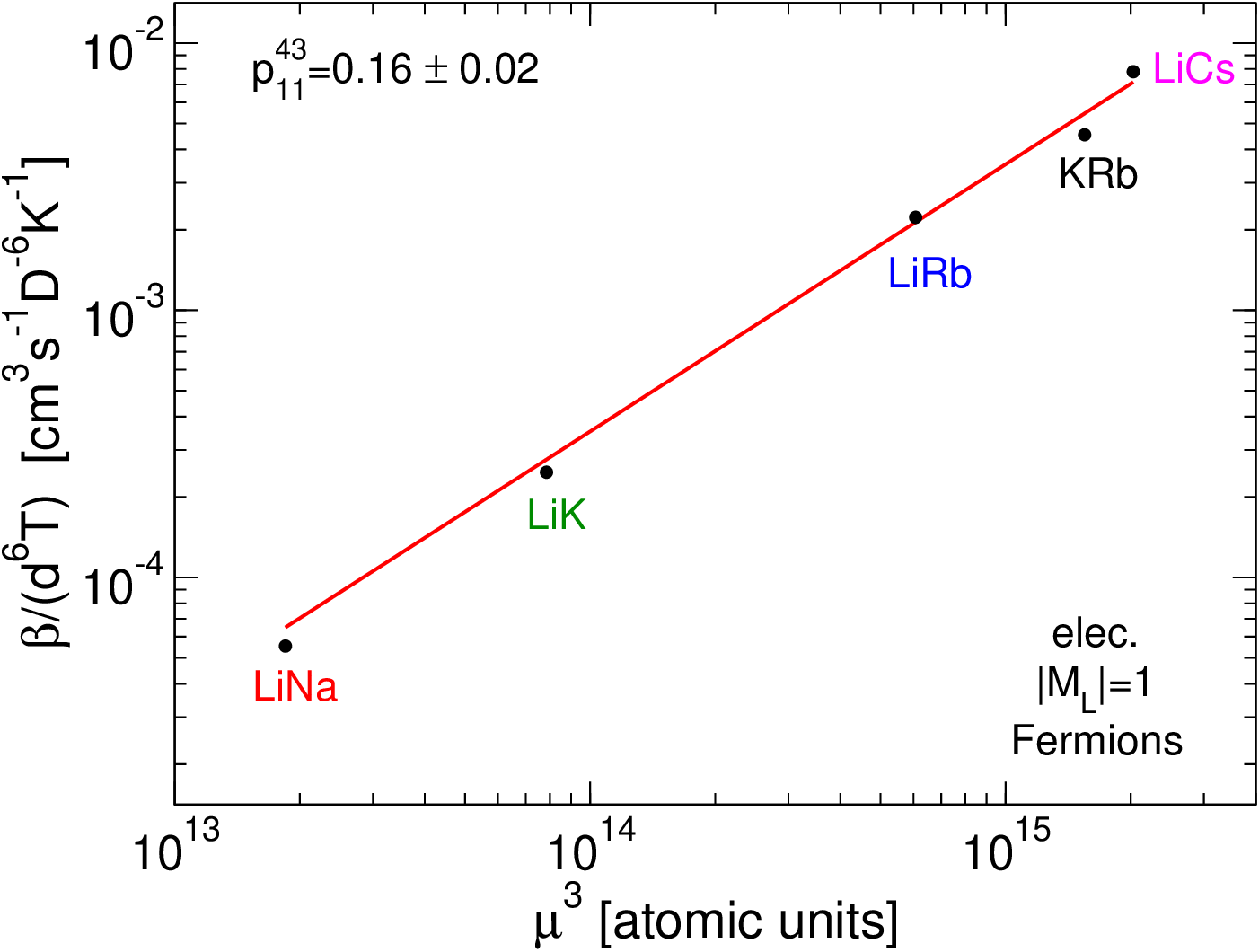}
\caption{(Color online) 
Top panel: Van der Waals regime for $M_L=0$ and $|M_L|=1$. 
The quantity $\beta_{1,0}^{6,2} = \beta_{1,1}^{6,2} $ divided by $T$ is plotted as a function of $(\mu \, C_6)^{3/4}$.
Middle panel: Electric field regime for $M_L=0$.
The quantity $\beta_{1,0}^{3,2}$ divided by $d^6 \, T$ is plotted as a function of $\mu^{3}$.
Bottom panel: Electric field regime for $|M_L|=1$.
The quantity $\beta_{1,1}^{4,3}$ divided $ d^6 \, T$ is plotted as a function of $\mu^{3}$.
\label{FER3D-PART-FIG}
}
\end{center}
\end{figure}

Replacing these three values of $E_*$ 
into Eq.\eqref{rateQT} for $L=1$ and assuming $\langle E_c \rangle = 3 k_B T / 2$, where $k_B$
is the Boltzmann constant and $T$ the temperature, 
we arrive at the following expressions for the $|M_L|=0,1$ rate as $d \to 0$ in the van der Waals regime
\begin{multline}
\beta_{L=1, |M_L|=0,1}^\text{ vdW} 
= \\
p_{1,(0,1)}^{6,2} \ \frac{\pi}{8} \,  
\left( \frac{3^{13} \, \mu^3 \, C_6^3}{\hbar^{10}} \right)^{1/4}  \, k_B T \times \Delta.
\label{rateQT10or162}
\end{multline}
with
\begin{eqnarray}
p_{1,(0,1)}^{6,2} = 0.53 \pm 0.07.
\label{pQT10or162}
\end{eqnarray}
The $|M_L|=0$ rate as $d \to d_\text{p}$ in the electric field regime is
\begin{multline}
\beta_{L=1, |M_L|=0}^\text{ elec}  
 = 
p_{1,0}^{3,2} \ \frac{3 \pi}{8} \, 
\left( \frac{6^9}{5^6} \right)^{1/2} \, \frac{\mu^3}{\hbar^{7}} \\
\, \frac{d^{6}}{(4 \pi \varepsilon_0)^{3}}
\, k_B T \times \Delta 
\label{rateQT1032}
\end{multline}
with
\begin{eqnarray}
p_{1,0}^{3,2} = 0.54 \pm 0.04.
\label{pQT1032}
\end{eqnarray}
Finally, the $|M_L|=1$ rate as $d \to d_\text{p}$ in the electric field regime is given by
\begin{multline}
\beta_{L=1, |M_L|=1}^\text{ elec}  
 = 
p_{1,1}^{4,3} \ \frac{3 \pi}{8} \, 
\left( 20000 \, (24/875)^3 \right)^{3/2} \, \frac{\mu^3}{\hbar^{7}}  \\
\, \frac{d^{6}}{(4 \pi \varepsilon_0)^{3}}
\, k_B T \times \Delta 
\label{rateQT1143}
\end{multline}
with
\begin{eqnarray}
p_{1,1}^{4,3} = 0.16 \pm 0.02.
\label{pQT1143}
\end{eqnarray}
The coefficients $p_{L,|M_L|}^{n,m}$ associated with the characteristic energies $E_{_{L,|M_L|}}^{n,m}$,
are found by confronting the analytical results in Eq.~\eqref{rateQT10or162}, Eq.~\eqref{rateQT1032}, and Eq.~\eqref{rateQT1143}
with our numerical calculations of Fig.~\ref{FER3D-TOTAL-FIG}.
The quantity $\beta_{L=1,|M_L|=0,1}^\text{ vdW}$ divided by $T$ obtained from the numerical results, 
is plotted as a function of the quantity $(\mu \, C_6)^{3/4}$ for the van der Waals 
regime in the top panel of Fig.~\ref{FER3D-PART-FIG}.
The quantities $\beta_{L=1,|M_L|=0}^\text{ elec}$ and $\beta_{L=1,|M_L|=1}^\text{ elec}$ divided by $d^6$ and $T$ 
are plotted as a function of the quantity 
$\mu^3$ for the electric field regime for the $|M_L=0|$ and $|M_L=1|$ component 
in the middle and bottom panels of Fig.~\ref{FER3D-PART-FIG} respectively,
for the different fermionic reactive systems.
We find that the numerical results fit a line, confirming the validity of the QT model analysis
(the fitting uncertainty of the lines provides an uncertainty to the $p_{L,|M_L|}^{n,m}$ parameters).
The fitting parameters are the slope of these lines and 
are reported in Eq.~\eqref{pQT10or162}, Eq.~\eqref{pQT1032} and Eq.~\eqref{pQT1143}.

We see that both components $|M_L|=0,1$ analytical rates (Eq.\eqref{rateQT10or162})
at ultracold temperature are the same in the van der Waals regime and 
are dictated by a $d^6$ dependence in the electric regime (Eq.\eqref{rateQT1032} and Eq.~\eqref{rateQT1143}) with different magnitudes.
These expressions provide a clear explanation of the trends observed numerically. 
The loss rate behaves as $(\mu \, C_6)^{3/4}$ in the van der Waals regime.
In the electric field regime, the loss rate scales 
as $\mu^3$, increasing only with the mass.
In both regimes, these expressions explain why fermionic LiNa is
the least reactive alkali polar species and fermionic LiCs is the most reactive one. 

We note that the results of $p_{1,(0,1)}^{6,2} = 0.53 \pm 0.07$ is 
in very good agreement with the analytical expression of $2^{19/4} \, \pi / (3^{17/4} \, [\Gamma(3/4)]^2) = 0.528$
found using a Quantum Defect Theory (QDT)~\cite{Idziaszek10-PRL}.
The values $p_{1,0}^{3,2} = 0.54 \pm 0.04$ and $p_{1,1}^{4,3} = 0.16 \pm 0.02$
for the $1/R^3$ interaction in the electric field regime
have not to our knowledge been determined analytically in a QDT framework.

We also note that these constants barely change
between the regime dominated by the van der Waals interaction
and the regime dominated by an electric field interaction for the $|M_L|=0$ component.
The ratio of the $|M_L|=1$ over the $|M_L|=0$ component in the electric field regime
is 0.003. As a consequence, the $|M_L|=1$ component is negligible in the electric regime, 
as seen in Fig.~\ref{FER3D-TOTAL-FIG} for the LiNa system,
and one can provide 
an estimation of the total p-wave rate coefficient
for the reactive systems by
\begin{eqnarray}
\beta_{L=1} &=& \beta_{L=1, |M_L|=0} + 2 \, \beta_{L=1, |M_L|=1}  \nonumber \\
 & \approx & 3 \, \beta_{L=1, |M_L|=0,1}^\text{ vdW} + \beta_{L=1, |M_L|=0}^\text{ elec} \nonumber \\
 & \approx & \frac{\pi}{8} \,  
\bigg\{ 
0.53 \times
\left( \frac{3^{17} \, \mu^3 \, C_6^3}{\hbar^{10}} \right)^{1/4} \nonumber \\
& + & 0.54 \times
\, \left( \frac{2^{9/2} \, 3^{11/2} \, \mu^3} {5^3 \, \hbar^{7}} \right)
\, \frac{d^{6}}{(4 \pi \varepsilon_0)^{3}}
\bigg\}  
\, k_B T \times \Delta.
\end{eqnarray}

\subsubsection{QT model for s-wave collisions}

For s-wave collisions($L=0, M_L=0$),
there is no incident barrier
because the repulsive centrifugal term vanishes.
It is possible however to estimate 
a characteristic length and energy~\cite{Gao08}
given respectively by
\begin{eqnarray}
a_n = \left( \frac{2 \, \mu \, C_n}{\hbar^2} \right)^{\frac{1}{n-2}} \quad ; \quad
E^n_{_{L=0,M_L=0}} = \frac{\hbar^2}{2 \, \mu \, a_n^{2} } .
\end{eqnarray}
In the van der Waals regime,
the characteristic energy is 
$E^6_{{0,0}}=\hbar^3 /  \sqrt{2^3 \, \mu^3 \, C_6}$. 
In the electric field regime,
the electric dipole-dipole interaction vanishes for $L=0$.
But as there is a coupling between the $L=0$ and the $L=2$ 
component in Eq.~\eqref{C3},
it is found after diagonalisation,
that the electric dipole interaction behaves as a $-C_4 / R^4$ 
with $C_4=4 \, \mu \, d^4/ [15 \, \hbar^2 \, (4 \pi \varepsilon_0)^2]$ (see Appendix C).
In return, this corresponds 
to a characteristic energy 
$E^4_{{0,0}}=15 \, \hbar^6 \, (4 \pi \varepsilon_0)^2/[16 \, \mu^3 \, d^4]$.
This is summarized in Tab~\ref{TAB3}.

\begin{table}[h]
\caption{Characteristic energies $E_*$ for $L=0, |M_L|=0$
in the van der Waals (vdW) and electric (elec.) regime.
\label{TAB3}
}
\begin{center}
\begin{tabular}{|c | c | c |c|}
\hline
regime & $|M_L|$ & - $C_n / R^n$ &  $E_{_{L=0,|M_L|}}^{n}$    \\ [0.5ex]
\hline
 vdW  &0 & - $\frac{C_6}{R^6}$ & $\frac{\hbar^3}{\sqrt{2^3 \, \mu^3 \, C_6}} $ \\
 elec. &0 & - $\frac{ 4 \, \mu \, d^4}{15 \, \hbar^2 \, (4 \pi \varepsilon_0)^2 \, R^4}$ & $\frac{15 \, \hbar^6}{16 \, \mu^3} \left( \frac{d^2}{4 \pi \varepsilon_0} \right)^{-2}$ \\ [0.5ex]
\hline
\end{tabular}
\end{center}
\end{table}

\begin{figure} [t]
\begin{center}
\includegraphics*[width=7.5cm,keepaspectratio=true,angle=0]{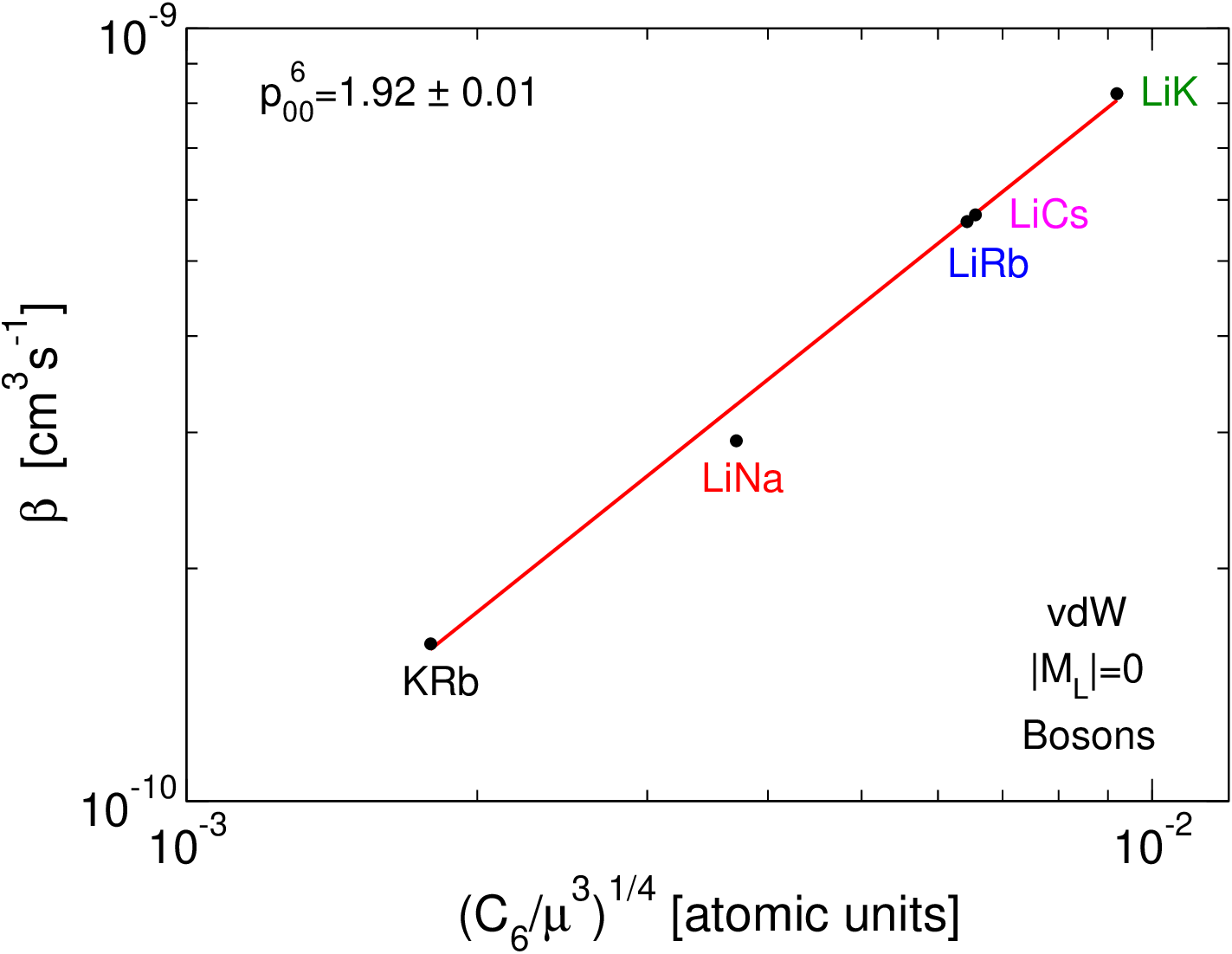}  \\
\includegraphics*[width=7.5cm,keepaspectratio=true,angle=0]{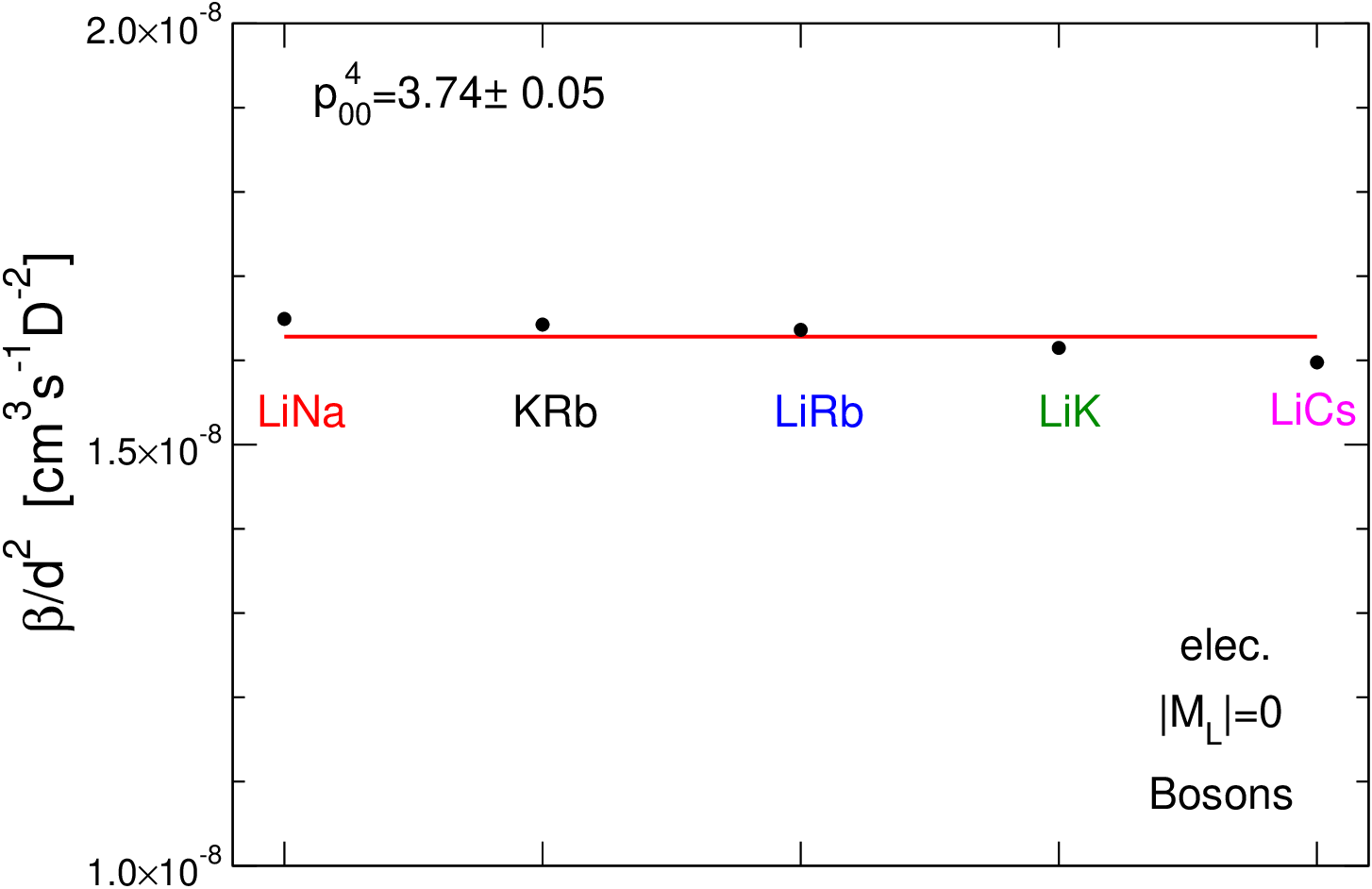}  
\caption{(Color online) 
Top panel: Van der Waals regime. 
The quantity $\beta_{0,0}^{6}$ is plotted as a function of $(C_6 / \mu^3)^{1/4}$.
Bottom panel: Electric field regime.
The quantity $\beta_{0,0}^{4}$ divided by $d^{2}$ is plotted for the five different colliding species.
\label{BOS3D-PART-FIG}
}
\end{center}
\end{figure}

Replacing these two values of $E_*$ 
into Eq.\eqref{rateQT} for $L=0$,
we arrive at the following expression for the $L=0, |M_L|=0$ rate as $d \to 0$ in the van der Waals regime
\begin{eqnarray}
\beta_{L=0, |M_L|=0}^\text{ vdW} 
= p_{0,0}^{6} \ 
\pi \, \left( \frac{2 \, \hbar^{2} \, C_6}{\mu^{3}} \right)^{1/4}   \times \Delta
\label{rateQT006}
\end{eqnarray}
with
\begin{eqnarray}
p_{0,0}^{6} = 1.92 \pm 0.01.
\label{pQT006}
\end{eqnarray}
The $L=0,|M_L|=0$ rate as $d \to d_\text{p}$ in the electric field regime is
\begin{eqnarray}
\beta_{L=0, |M_L|=0}^\text{ elec}  
 = p_{0,0}^{4} \ 
\pi \, \frac{\sqrt{16/30}}{\hbar} \, \frac{d^{2}}{4 \pi \varepsilon_0} \times \Delta 
\label{rateQT004}
\end{eqnarray}
with
\begin{eqnarray}
p_{0,0}^{4} = 3.74 \pm 0.05.
\label{pQT004}
\end{eqnarray}
Compared to $L=1$, the rates at ultracold temperature for $L=0$ behave now as $(C_6 / \mu^3)^{1/4}$ in the van der Waals regime,
making bosonic KRb molecules the least reactive ones and bosonic LiK molecules the most reactive ones,
due to the interplay between the $C_6$ coefficients and the cube of the mass.
In the electric field regime, the rates behave as $d^2$ and are independent of the mass, so that
for the same induced dipole, all bosonic polar molecules react with the same rate coefficient.
The coefficients $p_{L,|M_L|}^{n}$ associated with the characteristic energies $E_{_{L,|M_L|}}^{n}$,
are found by plotting the quantity $\beta_{L=0,|M_L|=0}^\text{ vdW}$ obtained from the numerical results of Fig.~\ref{BOS3D-TOTAL-FIG}
as a function of the quantity $(C_6 / \mu^3)^{1/4}$ for the van der Waals 
regime in the top panel of Fig.~\ref{BOS3D-PART-FIG},
and the quantity $\beta_{L=0,|M_L|=0}^\text{ elec}$ divided by $d^{2}$ 
for the electric regime in the bottom panel of Fig.~\ref{BOS3D-PART-FIG},
for the different bosonic reactive systems.
As for the fermionic case, the numerical results form a line for the first plot and are constant for the second plot,
validating the QT model analysis.
Again, we note that the results of $p_{0,0}^{6} = 1.92 \pm 0.01$ in Eq.~\eqref{pQT006} is 
in very good agreement with the analytical expression of $8 \, \pi / [\Gamma(1/4)]^2 = 1.912$
found using a Quantum Defect Theory~\cite{Idziaszek10-PRL} or a Quantum Langevin     
Theory (QL)~\cite{Gao10}.
The value of $p_{0,0}^{4} = 3.74 \pm 0.05$  agrees within 7\% with the analytical expression
of 4
from a Quantum Langevin
Theory~\cite{Gao11} using the $-C_4/R^4$ interaction in the electric field regime.
One can formulate a good approximation for the s-wave loss rate coefficients by
\begin{eqnarray}
\beta_{L=0} &=& \beta_{L=0, |M_L|=0}^\text{ vdW} + \beta_{L=0, |M_L|=0}^\text{ elec} \nonumber \\
 & \approx & \pi \, 
\bigg\{ 
 1.92 \times
 \left( \frac{2 \, \hbar^{2} \, C_6}{\mu^{3}} \right)^{1/4}  \nonumber \\
& + & 3.74 \times
 \frac{\sqrt{16/30}}{\hbar} \, \frac{d^{2}}{4 \pi \varepsilon_0} 
\bigg\}  
\times \Delta . 
\end{eqnarray}  \\

The formulas from Eq.~\eqref{rateQT10or162} to Eq.~\eqref{pQT1143} 
and from Eq.~\eqref{rateQT006} to Eq.~\eqref{pQT004} can be used to determine
the inelastic and reactive collisional properties of other atom-atom,
atom-molecule or molecule-molecule collisions, provided that full loss occurs
when they encounter one another.
This case can occur for molecules of category 2 (NaK, NaRb, NaCs, KCs, RbCs)
if the molecules are not in their absolute ground state,
for example in a higher vibrational state, where inelastic molecule-molecule collision can occur
or when the reactants have higher energy than the products so that
an exoergic reaction can take place. 
What is left unknown is the $C_6$ coefficients (except for RbCs), 
for each of these initial ro-vibrational states
of these molecules and has to be calculated individually.
For the RbCs molecule the $C_6$ coefficients have been calculated 
as a function of vibrational quantum number in Ref.~\cite{Kotochigova10}. \\

We provide in Appendix D the corresponding QT expressions 
for the imaginary part of the scattering lengths for s-wave collisions
and scattering volumes for p-wave collisions.

\section{Dynamics in two-dimensional space}

\begin{figure} [t]
\begin{center}
\includegraphics*[width=7.5cm,keepaspectratio=true,angle=0]{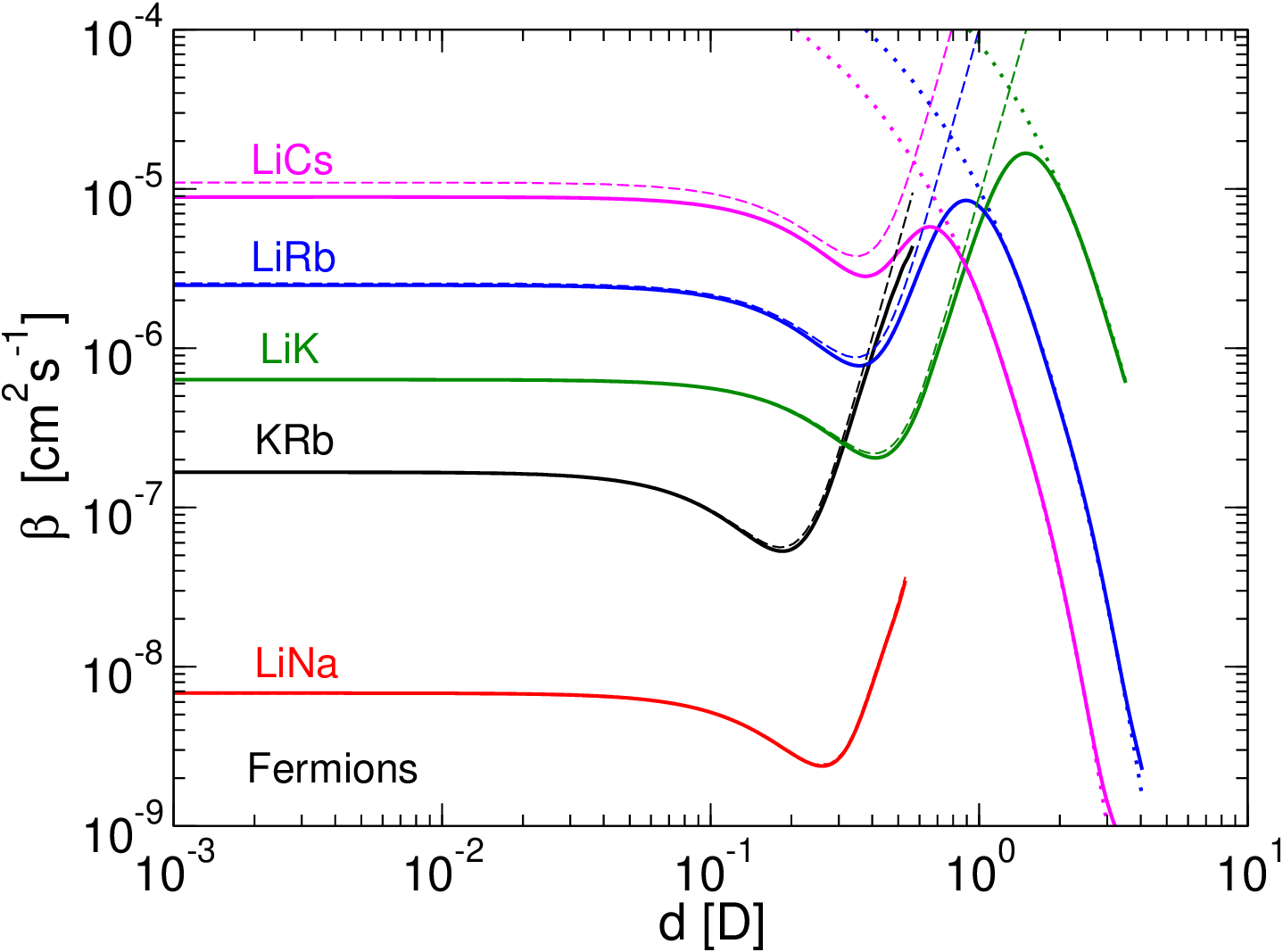}  \\
\includegraphics*[width=7.5cm,keepaspectratio=true,angle=0]{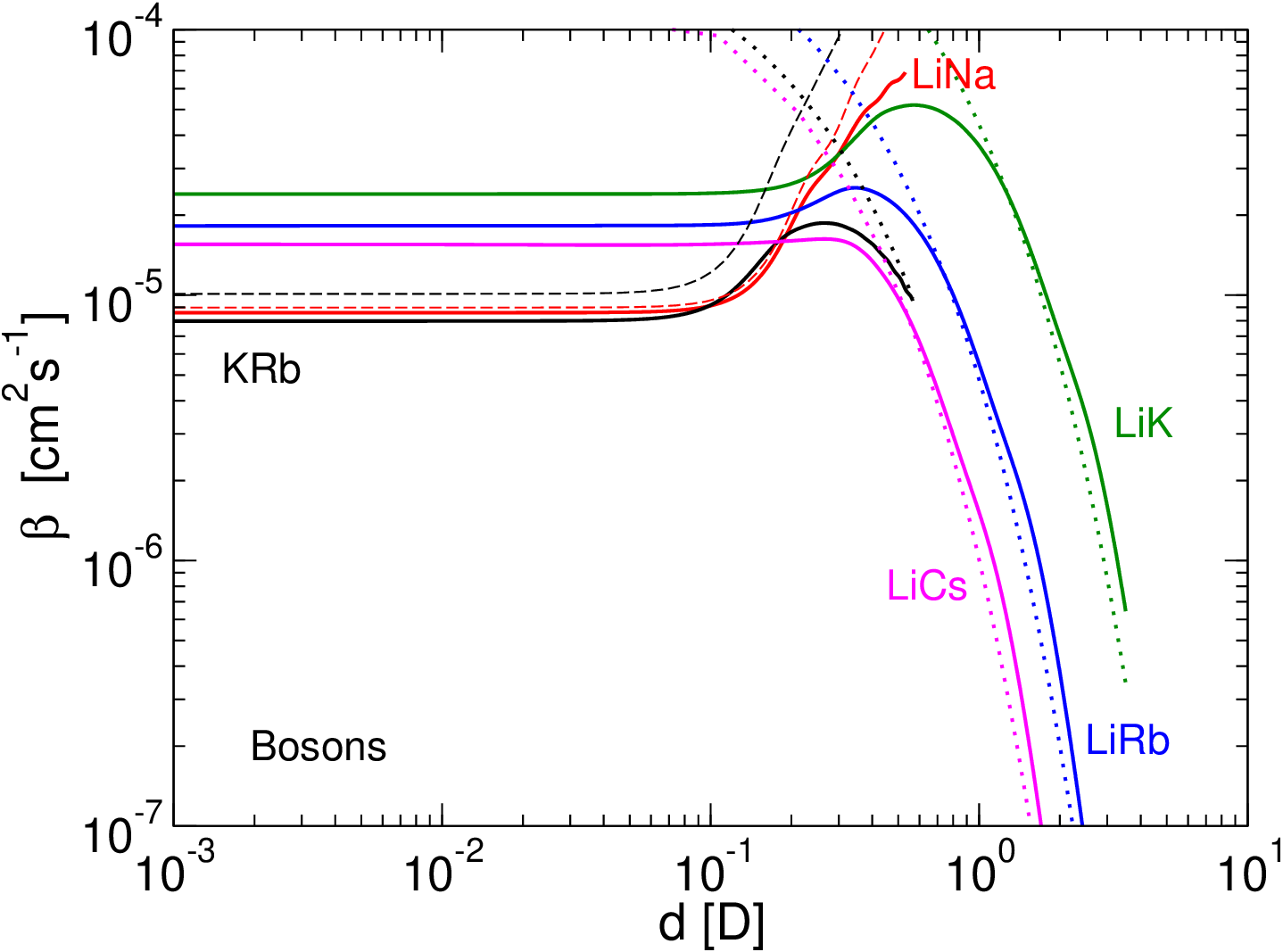} 
\caption{(Color online) 
Loss rate coefficients for different reactive polar molecules in confined 2D space,
for indistinguishable fermions (top panel) and indistinguishable bosons (bottom panel).
The frequency of the 1D trap is $\nu=20$~kHz and the temperature is $T=500$~nK.
The dashed lines represent a model based on the rescaled 3D rate coefficients for $a_\text{dd} \ll a_\text{ho}$,
and the dotted lines represent a model based on a functional form (Ref.~\cite{Julienne11-PCCP} for example) 
for $a_\text{dd} \gg a_\text{ho}$.
\label{2D-FIG}
}
\end{center}
\end{figure}

For the confined 2D scattering we use the same 
formalism developed in Ref.~\cite{Quemener10-2D,Quemener11-FULL}. 
The confinement is given by an optical lattice in the $\hat{z}$ direction,
which we approximate by a harmonic oscillator potential $V_\text{ho} = \mu \omega^2 z^2 / 2$ 
of frequency $\nu$ and angular frequency $\omega = 2 \, \pi \, \nu$.
One can also define a harmonic oscillator 
confinement length $a_\text{ho} = \sqrt{\hbar/(\mu \, \omega)}$.
We consider the dynamics of two molecules in the ground state
of this harmonic oscillator.
In confined space, $M_L$ remains a good quantum number.
Additional selection rules apply and
for indistinguishable bosons, $|M_L|=0$,
while for indistinguishable fermions, $|M_L|=1$~\cite{Quemener10-2D,Quemener11-FULL}, 
for molecules in the ground state of the harmonic confinement.
We present in Fig.~\ref{2D-FIG} the loss rate coefficient for a 
confinement of $\nu = 20$~kHz as a function of the dipole moment
for a given temperature $T = 500$~nK, for the fermionic
species (top panel) and the bosonic species (bottom panel).
We use fourty  partial waves, $L=1-79$ for the fermions
and $L=0-78$ for the bosons,
to converge the results.
At small electric dipoles, when $a_\text{dd} \ll a_\text{ho}$,
the collisions are quasi-2D (q2D) and the loss rate coefficients
display a similar behavior than their 3D counterpart 
for $|M_L|=1$ for the indistinguishable fermions
and for $|M_L|=0$ for the indistinguishable bosons.
At large electric dipoles and for LiK, LiRb, LiCs,  
when $a_\text{dd} \gg a_\text{ho}$,
the collisions are fully 2D and the loss rate coefficients
show a suppression as discussed in 
Ref.~\cite{Ticknor10,Quemener10-2D,Micheli10-PRL,Idziaszek10-RAPID,Quemener11-FULL,Dincao10,Julienne11-PCCP}.

For the quasi-2D regime $a_\text{dd} \ll a_\text{ho}$, we compare in dashed lines in Fig.~\ref{2D-FIG}
a two dimensional loss rate coefficient rescaled from the numerical calculation in three dimensions~\cite{Petrov01,Li09,Micheli10-PRL}
from the previous section
for $L=1$
\begin{eqnarray}
\beta^{q2D} = \frac{3}{2} \, \frac{\beta^{3D}}{\sqrt{\pi} \, a_\text{ho}} = \frac{3}{2} \, \sqrt{\frac{\mu \, \omega}{\pi \, \hbar}} \, \beta^{3D}
\label{QTq2D-F}
\end{eqnarray}
where the factor 3/2 accounts for the difference of the mean energies in 3D and 2D 
for a given temperature $T$ (in 3D, $\langle E_c \rangle = 3 k_B T /2$
while in 2D, $\langle E_c \rangle = k_B T $).
For $L=0$, we get
\begin{eqnarray}
\beta^{q2D} = \frac{\beta^{3D}}{\sqrt{\pi} \, a_\text{ho}} = \sqrt{\frac{\mu \, \omega}{\pi \, \hbar}} \, \beta^{3D}.
\label{QTq2D-B}
\end{eqnarray}
We found for the fermions a good agreement between the numerical 2D rates (in solid lines) 
and the rescaled from 3D rates (dashed lines).
For the bosons, a good agreement is found for the LiNa system, but not for the other systems like KRb for example,
even if the order of magnitude is right.
For bosons, threshold laws display a logarithmic dependence and are not accounted in Eq.~\eqref{QTq2D-B}.
The QT formulas which describe the numerical 3D rates can also be rescaled in the same manner so that
a good approximation to the loss rate coefficient for fermions in the quasi-2D regime $a_\text{dd} \ll a_\text{ho}$ is given by
\begin{eqnarray}
\beta^{q2D}_{L=1} & = & 2 \times \frac{3}{2} \, \sqrt{\frac{\mu \, \omega}{\pi \, \hbar}} \, \beta^{3D}_{|M_L|=1} \nonumber \\
 & \approx & 3 \sqrt{\frac{\mu \, \omega}{\pi \, \hbar}} \, 
\bigg\{ 
0.53 \times
\left( \frac{3^{13} \, \mu^3 \, C_6^3}{\hbar^{10}} \right)^{1/4} \nonumber \\
 & + & 0.16 \times
\frac{3 \pi}{8} \, 
\left( \frac{20000}{27} \frac{72}{875}\right)^{3/2} \, \frac{\mu^3}{\hbar^{7}}  \nonumber \\
& & \frac{d^{6}}{(4 \pi \varepsilon_0)^{3}}
\bigg\}
\, k_B T \times \Delta.
\end{eqnarray}
For bosons, the rescaled QT formula is 
\begin{eqnarray}
\beta^{q2D}_{L=0} & = &  \sqrt{\frac{\mu \, \omega}{\pi \, \hbar}} \, \beta^{3D}_{|M_L|=0} \nonumber \\
 & \approx &  \sqrt{ \frac{\pi \, \mu \, \omega}{\hbar}} \, 
\bigg\{ 
 1.92 \times 
\left( \frac{2 \, \hbar^{2} \, C_6}{\mu^{3}} \right)^{1/4}  \nonumber \\
& + & 3.74 \times 
 \frac{\sqrt{16/30}}{\hbar} \, \frac{d^{2}}{4 \pi \varepsilon_0} 
\bigg\}  
\times \Delta 
\end{eqnarray}
but is only a good approximation for LiNa.

For the 2D regime $a_\text{dd} \gg a_\text{ho}$, we compare in Fig.~\ref{2D-FIG}
a functional form provided in Refs.~\cite{Julienne11-PCCP,Buchler07,Micheli07,Micheli10-PRL,Ticknor09,Ticknor10,Dincao10}.
We found that the forms
\begin{eqnarray}
\beta^{2D} = 2 \times 13 \, \frac{\hbar}{\mu} \left( \frac{E_c}{\hbar \, \omega/2}\right)^2 \, e^{-2 \, (a_\text{dd} / a_\text{ho})^{2/5}} \times \Delta 
\end{eqnarray}
for indistinguishable fermions, and
\begin{eqnarray}
\beta^{2D} = 13 \, \frac{\hbar}{\mu} \left( \frac{E_c}{\hbar \, \omega/2}\right)^2 \, e^{-2 \, (a_\text{dd} / a_\text{ho})^{2/5}} \times \Delta 
\end{eqnarray}
for indistinguishable bosons fit well the numerical data.
These formulas are reported in dotted lines in Fig.~\ref{2D-FIG}.
We find a coefficient of $13$ in front of the exponential
and a coefficient of $2$ inside the exponential,
by fitting our numerical results.
These values                                          
are different from the values found in Ref.~\cite{Julienne11-PCCP}.
This is attributed to the different regimes of collision energies and confinements
involved in the fitting. It has been shown in Ref.~\cite{Julienne11-PCCP}
that the fitting parameters of the functional form may differ 
for different values of the collision energies.

\section{Conclusion}

By computing the $C_6$ coefficients for different pairs of alkali polar molecules of LiNa, LiK, LiRb, and LiCs,
and using an available one for KRb,
we estimated the quenching rate coefficient assuming full loss when they encounter one another,
for the fermionic species and for the bosonic species, both for the van der Waals regime and the electric field regime. 
We found that, at ultracold temperature, fermionic LiNa is the least reactive system while 
LiCs is the most in the van der Waals regime and electric field regime,
due mainly to the increase of the $C_6$ coefficient for the former regime and due to the increase of the mass for the later.
Bosonic KRb molecules are found to be the least reactive ones while LiK the most in the van der Waals regime.
All the bosonic molecules were found to have the same universal reactive rate in the electric field regime.
These behaviors were all explained using a Quantum Threshold model.
From our numerical results, we found analytical expressions for the reactive rate coefficients
for fermionic and bosonic molecules, in the van der Waals and electric field regime.
These expressions can be used for other type of systems, such as atom-molecule or molecule-molecule collisions
assuming full inelastic or reactive loss, if the corresponding $C_6$ coefficients are known.
For example, the analytical expressions can be applied to collision of non-ground state molecules 
of NaK, NaRb, NaCs, KCs and RbCs.
The present study provides useful information about collisional properties 
of heteronuclear alkali polar molecules for which increasingly experimental interest is devoted.
Future studies will consider the vibrational and rotational dependence of the $C_6$ coefficient
of the heteronuclear alkali molecules,
the higher anisotropic terms in the long-range interaction,
as well as the effect of higher collision energies, when more partial waves dominate.

\section*{Acknowledgments}

This material is based upon work supported by the Air Force Office of Scientific Research
under the Multidisciplinary University Research Initiative Grant No. FA9550-09-1-0588.
A. P. and S. K. are also grateful
for funding from NSF Grant PHY-1005453.

\section*{Appendix A: Characteristics of the heteronuclear alkali molecules}

We provide in Table~\ref{TAB4} 
a summary of the characteristics 
of the fermionic and bosonic isotopes studied in this work.
Conversion factors from atomic units (a.u.) to
S.I. units are:
1~a.u. of mass is equal to 1822.89~a.m.u. (atomic mass unit),
1~a.u. of electric dipole moment is equal to 2.5417~D,
1~a.u. of $C_6$ is equal to 1~E$_{\text{h}}$a$^6_0$ with
1~E$_{\text{h}}$ (Hartree) equal to 4.35974394$\times$10$^{-18}$~J and
1~a$_0$ (Bohr radius) equal to 0.529177$\times$10$^{-10}$~m.

\begin{table}[h]
\caption{Fermionic (F) or bosonic (B) character, isotope, reduced molecule-molecule mass $\mu$ (in a.u.), 
$C_6$ coefficient (in a.u.)
and permanent electric dipole moment $d_\text{p}$ (in D)
for the different heteronuclear alkali molecules.
\label{TAB4}
}
\begin{center}
\begin{tabular}{| c | c | c | c | c |}
\hline
F/B & isotope & $\mu$ (a.u.) & $C_6$ (a.u.) & $d_\text{p}$ (D)  \\ [0.5ex]
\hline
F & $^{6}$Li$^{23}$Na  & 26436 & \multirow{2}{*}{3880} & \multirow{2}{*}{0.531} \\
B & $^{7}$Li$^{23}$Na  & 27349 & & \\
\hline
F & $^{40}$K$^{87}$Rb  & 115638 & \multirow{2}{*}{16133} & \multirow{2}{*}{0.566} \\
B & $^{41}$K$^{87}$Rb  & 116547 & &  \\
\hline
F &  $^{7}$Li$^{40}$K  & 42820 & \multirow{2}{*}{524000} & \multirow{2}{*}{3.513} \\
B &  $^{6}$Li$^{40}$K  & 41907 & & \\
\hline
F & $^{6}$Li$^{87}$Rb  & 84695 & \multirow{2}{*}{1070000} & \multirow{2}{*}{4.046} \\
B & $^{7}$Li$^{87}$Rb  & 85608 & & \\
\hline
F & $^{6}$Li$^{133}$Cs  & 126618 & \multirow{2}{*}{3840000} & \multirow{2}{*}{5.355} \\
B & $^{7}$Li$^{133}$Cs  & 127531 & & \\
\hline
\end{tabular}
\end{center}
\end{table}

\section*{Appendix B: Height of the adiabatic barrier for the $|M_L|=1$ component in electric field. Mixing $L=1$ and $L=3$.}

In this case, we have two diabatic effective potential curves
\begin{eqnarray}
V_{L=1}(R) = \frac{2 \, \hbar^2}{2 \mu R^2}  - \frac{C_6}{R^6} + \frac{(2/5) \, d^2}{4\pi\varepsilon_0 \, R^3} \nonumber \\
V_{L=3}(R) = \frac{12 \, \hbar^2}{2 \mu R^2}  - \frac{C_6}{R^6} - \frac{(2/5) \, d^2}{4\pi\varepsilon_0 \, R^3} \nonumber \\
\end{eqnarray}
and a coupling
\begin{eqnarray}
W(R) = \frac{(2 \sqrt{126} / 35) \, d^2}{4\pi\varepsilon_0 \, R^3}.
\end{eqnarray}
In the case of $|W| \ll |V_{L=3}-V_{L=1}|$,
the adiabatic effective potential curves are given 
after diagonalisation by 
\begin{eqnarray}
E_{\pm}(R) = V_{L=3/1}(R) \pm \frac{72 \, \mu \, d^4}{875 \, \hbar^2 \, (4\pi\varepsilon_0)^2 \, R^4}
\end{eqnarray}
and especially the lower one
\begin{eqnarray}
E_{-}(R) = \frac{2 \, \hbar^2}{2 \mu R^2}  - \frac{C_6}{R^6} + \frac{(2/5) \, d^2}{4\pi\varepsilon_0 \, R^3} - \frac{C_4}{R^4}
\label{eminus}
\end{eqnarray}
with
\begin{eqnarray}
C_4 = \frac{72 \, \mu \, d^4}{875 \, \hbar^2 \, (4\pi\varepsilon_0)^2}.
\end{eqnarray}
At large $d$, the most repulsive potential in Eq.~\eqref{eminus} is $[(2/5) \, d^2]/[4\pi\varepsilon_0 \, R^3]$
and the most attractive is $-C_4/R^4$ so that the height of the barrier is
\begin{eqnarray}
E_{_{L=1,|M_L|=1}}^{n=4,m=3} = \frac{ (875/24)^3 \, \hbar^6 }{ 10000 \, \mu^3} \left( \frac{d^2}{4 \pi \varepsilon_0} \right)^{-2}.
\end{eqnarray}

\section*{Appendix C: Adiabatic potential for the $|M_L|=0$ component in electric field. Mixing $L=0$ and $L=2$.}

Now we have the two diabatic effective potential curves
\begin{eqnarray}
V_{L=0}(R) &=& - \frac{C_6}{R^6} \nonumber \\
V_{L=2}(R) &=& \frac{6 \, \hbar^2}{2 \mu R^2}  - \frac{C_6}{R^6} - \frac{(4/7) \, d^2}{4\pi\varepsilon_0 \, R^3} \nonumber \\
\end{eqnarray}
and the coupling between them
\begin{eqnarray}
W(R) = - \frac{2 \, d^2}{\sqrt{5} \, 4 \pi \varepsilon_0 \, R^3}.
\end{eqnarray}
In the case of $|W| \ll |V_{L=2}-V_{L=0}|$,
the adiabatic effective potential curves are given 
after diagonalisation by 
\begin{eqnarray}
E_{\pm}(R) = V_{L=2/0}(R) \pm \frac{4 \, \mu \, d^4}{15 \, \hbar^2 \, (4 \pi \varepsilon_0)^2 \, R^4}
\end{eqnarray}
and especially the lower one
\begin{eqnarray}
E_{-}(R) = - \frac{C_6}{R^6} - \frac{C_4}{R^4}
\end{eqnarray}
with
\begin{eqnarray}
C_4 = \frac{4 \, \mu \, d^4}{15 \, \hbar^2 \, (4 \pi \varepsilon_0)^2}.
\end{eqnarray}

\section*{Appendix D: QT expression for imaginary scattering lengths and scattering volumes}

We provide here the analytical
QT expressions for imaginary scattering lengths
and imaginary scattering volumes.
If we define the scattering length
and the scattering volume (see Ref.~\cite{Bala97}) by
\begin{eqnarray}
a &=& a_r - i \, a_i=-\delta(k)/k \\
V &=& V_r - i \, V_i=-\delta(k)/k^3,
\end{eqnarray}
for vanishing wave-vectors $k \to 0$,
the loss rate can be written as
\begin{eqnarray} 
\beta_{L=0} &=& \bigg( 4 \, \hbar \, \pi \, a_i / \mu \bigg) \times \Delta \nonumber \\
\beta_{L=1,M_L} &=& \bigg( 4 \, \hbar \, \pi \, k^2 \, V_i / \mu \bigg) \times \Delta 
\end{eqnarray}
for one component $M_L$.
Similarly, the elastic rate is given by
\begin{eqnarray} 
\beta^{\text{el}}_{L=0} &=& \bigg( 4 \, \hbar \, \pi \, k \, |a|^2 / \mu \bigg) \times \Delta \nonumber \\
\beta^{\text{el}}_{L=1,M_L} &=& \bigg( 4 \, \hbar \, \pi \, k^5 \, |V|^2 / \mu \bigg) \times \Delta.
\end{eqnarray}
To get the corresponding cross sections, one has to divide the rates by the relative velocity $v=\hbar \, k / \mu$.
Identifying the loss rate with the QT model, one gets
the imaginary scattering length in the van der Waals regime
\begin{eqnarray}
a_i = 1.92 \times \bigg( \frac{\mu^{1/4} \, C_6^{1/4}}{2^{7/4} \, \hbar^{1/2}} \bigg) ,
\label{ai1}
\end{eqnarray}
the imaginary scattering length in the electric field regime
\begin{eqnarray} 
a_i = 3.74 \times \, \bigg(  \frac{\mu}{\hbar^2 \, \sqrt{30}} \bigg) \, \frac{d^{2}}{4 \pi \varepsilon_0} 
\label{ai2} 
\end{eqnarray}
the imaginary scattering volume in the van der Waals regime
\begin{eqnarray} 
V_i = 0.53 \times \bigg(  \frac{ 3^{9/4} \, \mu^{3/4} \, C_6^{3/4}}{32 \, \hbar^{3/2}} \bigg) ,
\label{Vi1}
\end{eqnarray}
and the imaginary scattering volume in the electric field regime
\begin{eqnarray} 
V_i = 0.54 \times  \bigg(  \frac{ 3^{9/2} \, \mu^3}{\hbar^6 \, 2^{1/2} \, 5^3} \bigg) 
\, \frac{d^{6}}{(4 \pi \varepsilon_0)^3} .
\label{Vi2} 
\end{eqnarray}
In the case of lossy collisions, the imaginary parts $a_i$ or $V_i$
contributes to the elastic part of the rates.
As a consequence, they provide a minimum value for the elastic rates
$\beta^{\text{el}}_{L=0} = (4 \, \hbar \, \pi \, k \, a_i^2 / \mu) \times \Delta $ for s-wave collisions
and $\beta^{\text{el}}_{L=1,M_L} = (4 \, \hbar \, \pi \, k^5 \, V_i^2 / \mu) \times \Delta$
for p-wave collisions.
In other words, lossy collisions imply non-zero elastic cross sections or rate coefficients.


\begin{thebibliography}{65}

\bibitem{Sage05}
J. M. Sage,
S. Sainis, T. Bergeman, and D. DeMille,
Phys. Rev. Lett. {\bf 94}, 203001 (2005).

\bibitem{Hudson08}
E. R. Hudson, N. B. Gilfoy, S. Kotochigova, J. M. Sage, and D. DeMille,
Phys. Rev. Lett. {\bf 100}, 203201 (2008).

\bibitem{Ni08}
K.-K. Ni, S. Ospelkaus,
M. H. G. de Miranda, A. Pe'er, B. Neyenhuis, J. J. Zirbel, S. Kotochigova,
P. S. Julienne, D. S. Jin, and J. Ye,
Science {\bf 322}, 231 (2008).

\bibitem{Deiglmayr08}
J. Deiglmayr,
A. Grochola, M. Repp, K. M\"ortlbauer,
C. Gl\"uck, J. Lange, O. Dulieu, R. Wester, and M. Weidem\"uller,
Phys. Rev. Lett. {\bf 101}, 133004 (2008).

\bibitem{Aikawa10}
K. Aikawa, D. Akamatsu, M. Hayashi, K. Oasa, J. Kobayashi, P. Naidon, T. Kishimoto, M. Ueda, and S. Inouye,
Phys. Rev. Lett. {\bf 105}, 203001 (2010).

\bibitem{Ospelkaus10-PRL}
S. Ospelkaus, K.-K. Ni, G. Qu\'em\'ener, B. Neyenhuis,
D. Wang, M. H. G. de Miranda, J. L. Bohn, J. Ye, and D. S. Jin,
Phys. Rev. Lett. {\bf 104}, 030402 (2010).

\bibitem{Ni10-NATURE}
K.-K. Ni, S. Ospelkaus, D. Wang, G. Qu\'em\'ener, B. Neyenhuis, M. H. G. de Miranda,
J. L. Bohn, J. Ye, and D. S. Jin,
Nature,  {\bf 464} 1324 (2010).

\bibitem{Miranda11}
M. H. G. de Miranda, A. Chotia, B. Neyenhuis, D. Wang, G. Qu\'em\'ener, S. Ospelkaus,
J. L. Bohn, J. Ye, and D. S. Jin, 
Nature Physics, {\bf 7}, 502 (2011). 

\bibitem{Aymar05}
M. Aymar and O. Dulieu,
J. Chem. Phys. {\bf 122}, 204302 (2005).

\bibitem{Deiglmayr08-JCP}
J. Deiglmayr, M. Aymar, R. Wester, M. Weidem\"uller, and Olivier Dulieu,
J. Chem. Phys. {\bf 129}, 064309 (2008).

\bibitem{Zuchowski10}
P. S. $\dot{\text{Z}}$uchowski and J. M. Hutson,
Phys. Rev. A {\bf 81}, 060703(R) (2010).

\bibitem{Byrd10}
J. N. Byrd, J. A. Montgomery Jr., and R. C\^ot\'e,
Phys. Rev. A {\bf 82}, 010502(R) (2010).

\bibitem{Meyer10}
E. R. Meyer and J. L. Bohn,
Phys. Rev. A {\bf 82}, 042707 (2010). 

\bibitem{Meyer11}
E. R. Meyer and J. L. Bohn,
Phys. Rev. A {\bf 83}, 032714 (2011).

\bibitem{Ospelkaus10-SCIENCE}
S. Ospelkaus, K.-K. Ni, D. Wang, M. H. G. de Miranda, B. Neyenhuis, G. Qu\'em\'ener,
P. S. Julienne, J. L. Bohn, D. S. Jin, and J. Ye,
Science {\bf 327}, 853 (2010).

\bibitem{Quemener10-QT}
G. Qu\'em\'ener and J. L. Bohn,
Phys. Rev. A {\bf 81}, 022702 (2010).

\bibitem{Ticknor10-3B}
C. Ticknor and S. T. Rittenhouse,
Phys. Rev. Lett. {\bf 105}, 013201 (2010).

\bibitem{Wang11-BOS}
Y. Wang, J. P. D'Incao, C. H. Greene,
Phys. Rev. Lett. {\bf 106}, 233201 (2011).

\bibitem{Wang11-FER}
Y. Wang, J. P. D'Incao, C. H. Greene,
e-print arXiv:1106.6133.

\bibitem{Carr09}
L. D. Carr, D. DeMille, R. V. Krems, and J. Ye,
New J. Phys. {\bf 11}, 055049 (2009).

\bibitem{Micheli06-NATURE}
A. Micheli, G. K. Brennen and P. Zoller,
Nat. Phys. {\bf 2}, 341 (2006).

\bibitem{Pupillo-Chapter}
G. Pupillo, A. Micheli, H.-P. B\"uchler, and P. Zoller, in Cold
Molecules: Theory, Experiment, Applications, edited by R. V.
Krems, W. C. Stwalley, and B. Friedrich (CRC Press, Boca
Raton, FL, 2009).

\bibitem{Demille02}
D. DeMille,
Phys. Rev. Lett. {\bf 88}, 067901 (2002).

\bibitem{Yelin06}
S. F. Yelin, K. Kirby, and R. C\^ot\'e,
Phys. Rev. A {\bf 74}, 050301(R) (2006).

\bibitem{Gorshkov11-PRL}
A. V. Gorshkov, S. R. Manmana, G. Chen, J. Ye, E. Demler, M. D. Lukin, A.-M. Rey,
e-print arXiv:1106.1644.

\bibitem{Gorshkov11-PRA}
A. V. Gorshkov, S. R. Manmana, G. Chen, E. Demler, M. D. Lukin, A.-M. Rey,
e-print arXiv:1106.1655.

\bibitem{Ospelkaus08}
S. Ospelkaus, A. Pe'er, K.-K. Ni, J. J. Zirbel, B. Neyenhuis, S. Kotochigova, 
P. S. Julienne, J. Ye and D. S.Jin,
Nature Physics {\bf 4}, 622 (2008).

\bibitem{Kotochigova09}
S. Kotochigova, E. Tiesinga, and P. S. Julienne, 
New J. Phys. {\bf 11}, 055043 (2009).

\bibitem{Idziaszek10-PRL}
Z. Idziaszek and P. S. Julienne, 
Phys. Rev. Lett. {\bf 104}, 113202 (2010). 

\bibitem{Ticknor10}
C. Ticknor,
Phys. Rev. A {\bf 81}, 042708 (2010).

\bibitem{Kotochigova10}
S. Kotochigova, 
New J. Phys. {\bf 12}, 073041 (2010).

\bibitem{Quemener10-2D}
G. Qu\'em\'ener and J. L. Bohn,
Phys. Rev. A {\bf 81}, 060701(R) (2010).

\bibitem{Micheli10-PRL}
A. Micheli, Z. Idziaszek, G. Pupillo, M. A. Baranov, P. Zoller, and P. S. Julienne
Phys. Rev. Lett. {\bf 105}, 073202 (2010).

\bibitem{Idziaszek10-RAPID}
Z. Idziaszek, G. Qu\'em\'ener, J. L. Bohn, and P. S. Julienne
Phys. Rev. A {\bf 82}, 020703(R) (2010)

\bibitem{Gao10}
B. Gao,
Phys. Rev. Lett. {\bf 105}, 263203 (2010).

\bibitem{Quemener11-FULL}
G. Qu\'em\'ener and J. L. Bohn,
Phys. Rev. A {\bf 83}, 012705 (2011).

\bibitem{Dincao10}
J. P. D'Incao and C. H. Greene,
Phys. Rev. A {\bf 83}, 030702 (2011).

\bibitem{Julienne11-PCCP}
P. S. Julienne, T. M. Hanna, Z. Idziaszek,
Phys. Chem. Chem. Phys., 2011, Advance Article
DOI: 10.1039/C1CP21270B,
e-print arXiv:1106.0494.

\bibitem{Haimberger09}
C. Haimberger, J. Kleinert, P. Zabawa, A. Wakim, and N. P. Bigelow,
New. J. Phys. {\bf 11}, 055042 (2009).

\bibitem{Zabawa10}
P. Zabawa, A. Wakim, A. Neukirch, C. Haimberger, N. P. Bigelow, A. V. Stolyarov, E. A. Pazyuk, M. Tamanis, and R. Ferber,
Phys. Rev. A {\bf 82}, 040501(R) (2010).

\bibitem{Lercher11}
A. D. Lercher, T. Takekoshi, M. Debatin, B. Schuster, R. Rameshan, F. Ferlaino, R. Grimm, and H.-C. N\"agerl,
Eur. Phys. J. D, 2011, Advance article
DOI: 10.1140/epjd/e2011-20015-6,
e-print arXiv:1101.1409.

\bibitem{Debatin11}
M. Debatin, T. Takekoshi, R. Rameshan, L. Reichs\"ollner, F. Ferlaino, R. Grimm, R. Vexiau, N. Bouloufa, O. Dulieu, and H.-C. N\"agerl,
e-print arXiv:1106.0129.

\bibitem{Cho11}
H.W. Cho, D.J. McCarron, D. L. Jenkin, M. P. Köppinger, and S. L. Cornish,
Eur. Phys. J. D, 2011, Advance article
DOI: 10.1140/epjd/e2011-10716-1,
e-print arXiv:1107.5567.

\bibitem{Deiglmayr11-JPCS}
J. Deiglmayr, M. Repp, A. Grochola, O. Dulieu, R. Wester, and M. Weidem\"uller,
J. Phys.: Conf. Ser. {\bf 264} 012214 (2011).

\bibitem{Deiglmayr11-EPJD}
J. Deiglmayr, M. Repp, O. Dulieu, R. Wester, and M. Weidem\"uller,
e-print arXiv:1107.1060.

\bibitem{Ridinger11}
A. Ridinger, S. Chaudhuri, T. Salez, D. R. Fernandes, N. Bouloufa, O. Dulieu, C. Salomon, and F. Chevy,
e-print arXiv:1106.0494.

\bibitem{Stone}
A. J. Stone, 
{\it The theory of intermolecular forces}, (Clarendon Press,
London, 1996).

\bibitem{Tang}
K. T. Tang, 
Phys. Rev. {\bf 177}, 108 (1969).

\bibitem{Barnett}
R. Barnett, D. Petrov, M. Lukin, and E. Demler, 
Phys. Rev. Lett. {\bf 96}, 190401 (2006).

\bibitem{Herzberg}  
G. Herzberg, 
{\it Spectra of diatomic molecules}, 2$^{nd}$ edition (van Nostrand Company,
Princeton, 1950).

\bibitem{ccsd}
J. D. Watts, J. Gauss. and R. J. Bartlett, 
J. Chem. Phys. {\bf 98}, 8718 (1993).

\bibitem{Dunning:89} 
T.H. Dunning, Jr. J., 
Chem. Phys. {\bf 90}, 1007 (1989).

\bibitem{Prascher:11}  
B. P. Prascher, D. E. Woon,  K. A. Peterson, T. H., Jr. Dunning,
and A. K. Wilson, 
Theor. Chem. Acct. {\bf 128}, 69 (2011)

\bibitem{KPeterson2009} 
Kirk Peterson, 
private communication.

\bibitem{Stoll}
I. Lim, P. Schwerdtfeger, B. Metz, and H. Stoll, 
J. Chem. Phys. {\bf 122},
104103 (2005).

\bibitem{Bethe35}
H. A. Bethe,
Phys. Rev. {\bf 47}, 747 (1935).

\bibitem{Wigner48}
E. P. Wigner,
Phys. Rev. {\bf 73}, 1002 (1948).

\bibitem{Burke-PHD}
J. P. Burke Jr., Ph.D. thesis, University of Colorado (1999), available
online at http://jilawww.colorado.edu/pubs/thesis/burke.

\bibitem{Gao08}
B. Gao,
Phys. Rev. A {\bf 78}, 012702 (2008).

\bibitem{Gao11}
B. Gao,
Phys. Rev. A {\bf 83}, 062712 (2011).

\bibitem{Petrov01}
D. S. Petrov and G. V. Shlyapnikov,
Phys. Rev. A {\bf 64}, 012706 (2001).

\bibitem{Li09}
Z. Li and R. V. Krems,
Phys. Rev. A {\bf 79}, 050701(R) (2009).

\bibitem{Buchler07}
H. P. B\"uchler, E. Demler, M. Lukin, A. Micheli, N. Prokofiev, G. Pupillo, and P. Zoller,
Phys. Rev. Lett. {\bf 98}, 060404 (2007).

\bibitem{Micheli07}
A. Micheli, G. Pupillo, H. P. B\"uchler, and P. Zoller,
Phys. Rev. A {\bf 76}, 043604 (2007).

\bibitem{Ticknor09}
C. Ticknor,
Phys. Rev. A {\bf 80}, 052702 (2009).

\bibitem{Bala97}
N. Balakrishnan, V. Kharchenko, R. C. Forrey, and A. Dalgarno,
Chem. Phys. Lett {\bf 280}, 5 (1997).

\end{thebibliography}
\end{document}